\documentclass[runningheads]{llncs}
\usepackage[T1]{fontenc}
\usepackage{graphicx}
\usepackage{amsmath,amssymb}
\usepackage{booktabs}
\usepackage[hidelinks]{hyperref}
\graphicspath{{figures/}}

\newif\ifanon
\anonfalse % arXiv build; set \anontrue for ECIR submission

\begin{document}

\title{Closed Forms and Synthetic Twins: Predicting Approximate Nearest
Neighbor Recall from Embedding Statistics}
\titlerunning{Predicting ANN Recall from Embedding Statistics}

\ifanon
\author{Anonymous}
\institute{Anonymous institution}
\else
\author{Samuel Herman}
\institute{Independent Researcher\\
\email{sherman8915@gmail.com}\\
\url{https://sam-herman.github.io/}}
\fi

\maketitle

\begin{abstract}
Embedding models are trained and evaluated as if retrieval were exact;
in production they serve behind approximate indexes --- HNSW, IVF,
product quantization, or the fixed-dimensional encodings (FDEs) of
late-interaction models --- whose behavior the encoder's benchmarks never
see: one modern encoder recovers just 14\% of its exact top-10 through
its raw FDE index. Such failures surface only after an index is built,
and the standard patches --- corpus-fitted transforms such as whitening ---
must be fitted, stored, and refit as the corpus changes, and can
silently rewrite what the encoder returns. This paper shows that index
behavior is predictable before anything is built, from label-free statistics of
the raw embeddings, through a ladder of instruments matched to what
each index family consumes: (1) closed-form moment statistics for the
fixed-grid quantizers (PQ, FDE); (2) simulation on a synthetic twin
corpus --- cluster statistics made generative, on which any index,
composed production systems included, can be built and tested --- for
partition indexes; (3) size-extrapolated, lightly calibrated twins for
graph indexes at million-document scale. Predictions land within 0.03
of measured recall on an unseen million-document corpus. The same
geometry is trainable: targeting the one statistic no post-hoc
transform can move --- the score margin --- lifts recall for every index
family at once, at a small measured task cost. The result: index
choice, correction pricing, and production recall forecast from one
cheap measurement pass, on new corpora and new indexes alike; serving
without per-corpus transform machinery, suited to continuously changing
corpora; and a recall--compute frontier pushed by adapting encoders to
geometry rather than coupling them to any single index.

\end{abstract}

\section{Introduction}\label{sec:intro}

Embedding models are trained and evaluated as if retrieval were exact: the quality bar is computed on the top-$k$ of a brute-force nearest-neighbor search, and nothing in mainstream encoder development --- training, evaluation, or model selection --- asks what the serving index will do to that top-$k$ (a specialized co-design literature exists, Sect.~\ref{sec:related}; the encoders practitioners download are not trained that way). Consider a concrete case: GTE-ModernColBERT, a modern late-interaction encoder, reaches strong exact MaxSim quality on MS MARCO. At production scale, however, retrieval runs behind an approximation layer --- HNSW graphs~\cite{malkov2018hnsw}, IVF partitions, product quantization~\cite{jegou2011pq}, or the MUVERA-style fixed-dimensional encodings (FDEs)~\cite{dhulipala2024muvera} used to index multi-vector models --- and how faithfully that layer recovers the exact top-$k$ at a fixed compute budget depends strongly on measurable geometric properties of the embedding distribution. The same encoder's raw FDE recall@100 is $0.136$: the index recovers just $14\%$ of what the exact scorer would return, and nothing in a standard benchmark reveals it. Late-interaction FDEs are where the failure is most extreme, but it runs through every family of the approximation layer --- quantizers, IVF partitions, HNSW graphs --- and Sect.~\ref{sec:families} develops a predictive, quantitative account of each: closed forms where they hold, synthetic twins where they provably cannot.

\begin{figure}[t]
\includegraphics[width=\textwidth]{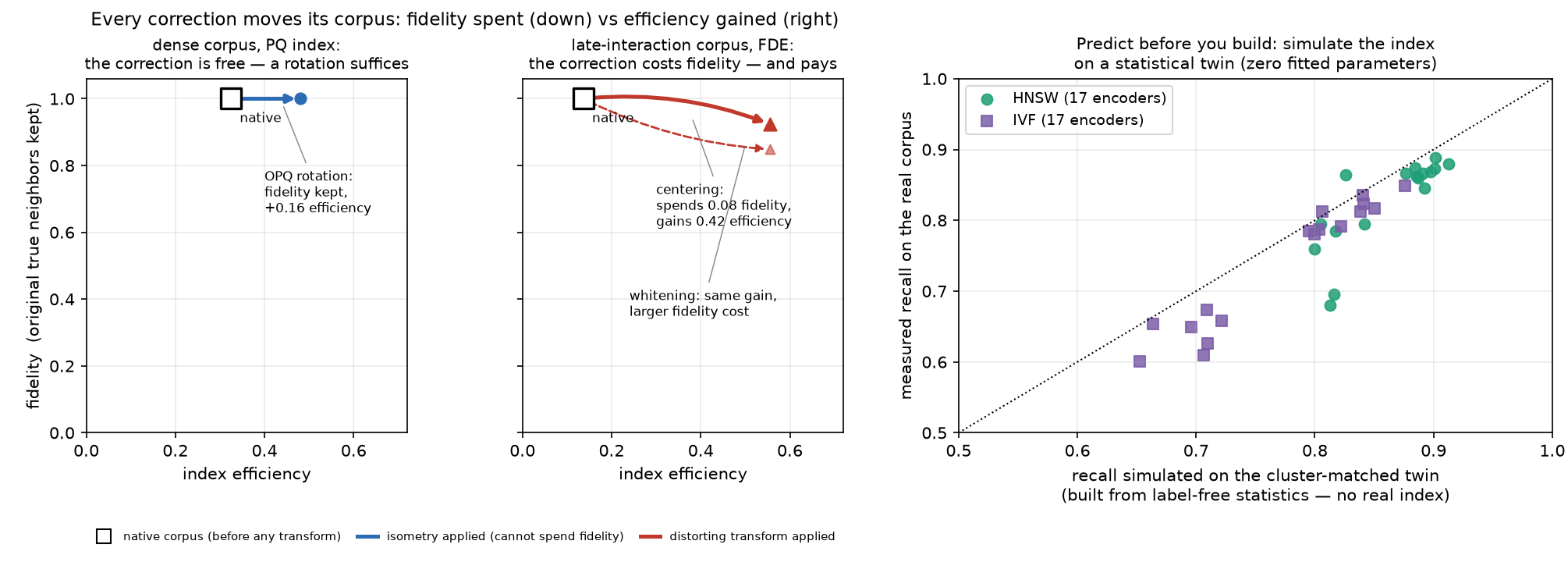}
\caption{The paper's two headline facts, from measured data. \emph{Left pair:} the fidelity/efficiency trade drawn as a before$\to$after move --- each panel starts at its corpus's native baseline (open square). On the dense corpus the correction is free: the OPQ rotation gains $0.16$ efficiency with fidelity kept. On the late-interaction corpus the correction costs fidelity and pays: centering spends $0.08$ for $0.42$ efficiency; whitening (dashed) buys the same gain for $0.15$, dominated by centering. Which trade is right depends on the index's tax (Sect.~\ref{sec:calculus} prices every measured transform; Fig.~\ref{fig:tradeoff} plots the full plane). \emph{Right:} recall simulated on a cluster-matched synthetic twin --- built purely from label-free statistics --- predicts measured recall for both navigation families across 17 encoders with zero fitted parameters ($25$--$50$k corpora; Sect.~\ref{sec:milliondoc} reports the million-document extension).}
\label{fig:teaser}
\end{figure}

Practitioners compensate with post-hoc, corpus-fitted transforms --- centering, whitening, learned rotations (OPQ~\cite{ge2013opq}; RaBitQ~\cite{gao2024rabitq}, now shipped in production systems) --- and they do make the space easier to index: centering more than triples GTE-ModernColBERT's FDE recall. But outside of pure rotations, which preserve every inner product and hence every exact ranking, the correction rewrites the exact neighbor structure itself --- after whitening BGE-large embeddings, $62\%$ of the true top-$10$ has changed (averaged over queries) before any index is built. Every corrective transform therefore trades two effects in direct tension: a semantic shift, paid up front, against the approximation error the index no longer makes; whether the trade pays depends on the index's tax --- the recall it loses to approximation (Sect.~\ref{sec:calculus}; Fig.~\ref{fig:teaser}, left). Index benchmarks cannot see the trade at all: they measure recall inside the transformed space, where the semantic shift is invisible by construction.

Neither effect is arbitrary. How much an index misses is set by the geometry of the point set it operates on; how much truth a transform rewrites is set by how far it moves points relative to one another. Both are mathematical consequences of measurable properties of the embedding distribution, so both should be predictable from measurements of it --- and prediction pays twice. It lets a practitioner reason about index choice, correction, and expected recall before building anything (Fig.~\ref{fig:teaser}, right). And once the governing properties are known and predictive, they become levers: the encoder can be trained to shape them at the source --- better served recall and, because no corpus-fitted correction remains in the serving path, indexes that absorb continuously changing corpora with no offline refitting step.

This paper therefore treats the geometry--index relationship as something to be \textbf{predicted, not narrated}, and asks three questions:

\begin{enumerate}
\item \emph{Prediction} --- which measurable properties of an embedding distribution determine each index family's behavior?
\item \emph{Pricing} --- when does a corrective transform pay for its semantic cost?
\item \emph{Learning} --- can the correction be learned with the model rather than fitted to the corpus?
\end{enumerate}

The prediction question's answer is a statistic-to-behavior \emph{map} --- the object the other two questions' payoffs depend on --- and it resolves at the level of index families: each \emph{family} --- not each individual index --- reads its own statistic of the embedding distribution. The FDE hashing grid reads the strength of the mean direction (the mean cosine similarity, MCS), product quantization the covariance's cross-block alignment (the misalignment factor), partitions the routing margin, and graph search the score margins over the corpus's cluster structure. These levers barely overlap, so geometry can be adapted for one family without disturbing another (Sects.~\ref{sec:families} and \ref{sec:learning}). For the navigation families --- partitions and graphs --- the governing cluster statistics stop short of a closed form; there, simulation on a cluster-matched twin carries the prediction. The paper adopts the simpler, cheaper instrument --- cluster-level statistics, nothing finer, parameter-free --- over a richer, scale-aware twin that would preserve fine-grained local structure as well, leaving the scale-aware construction beyond this paper's scope (Sect.~\ref{sec:limitations}); Sect.~\ref{sec:milliondoc} measures where the simple instrument reaches its limit and how to predict within it.

Throughout, the paper measures in two reference frames: fidelity to the base embedding's exact neighbors --- label-free, and sufficient on its own for judging indexes and corpus-fitted transforms --- and task relevance where labels exist, needed only where the correction is trained into the encoder itself and the label-free reference frame becomes the training target (Sect.~\ref{sec:groundtruth}). Within the label-free frame, the paper develops a predictive, end-to-end calculus: every transform is priced by the semantic cost it pays against the approximation gain it buys --- two effects no prior framework predicts jointly (Sects.~\ref{sec:axes} and \ref{sec:calculus}) --- and the per-family map identifies which side dominates for which index.

\paragraph{Contributions.} Each contribution answers one of the three questions, in order: prediction, pricing, learning.

\begin{enumerate}
\item \textbf{A predictive geometric analysis (Sects.~\ref{sec:framework}, \ref{sec:families} and \ref{sec:evaluation}).} A small set of label-free statistics --- the embedding mean and covariance read at the resolution each index consumes, plus cluster and margin statistics the moments provably do not determine --- explains and forecasts, per family, how geometry moves recall: a proven anisotropy-degeneration theorem for MUVERA-style FDEs exposing a hidden $(1-\mathrm{MCS})^{-1}$ factor in its data-oblivious guarantee (Theorem~\ref{thm:degeneration}, Corollary~\ref{cor:inflation}); a derived misalignment factor that attributes an encoder's PQ tax to its cause from the covariance alone; and a cluster-matched synthetic twin that predicts navigation recall by zero-parameter simulation, extending to million-document corpora with a single calibration measurement. Notably, on a held-out corpus of one million documents, the twin predicts index recall to within $0.03$. Every claim carries an explicit proved/derived/measured label (Sect.~\ref{sec:labels}).
\item \textbf{A trade-off calculus for corrective transforms (Sects.~\ref{sec:axes} and \ref{sec:calculus}).} Any transform's end-to-end effect decomposes into index efficiency gained versus exact-neighbor fidelity lost, with an exact lower bound on end-to-end recall in terms of the two; rotations pay zero fidelity and dominate whenever they suffice. The calculus predicts both observed real-data regimes: distortion is worth its price where the index tax is enormous (centering, on GTE-ModernColBERT's FDE), and silently loses where the tax is modest (on BGE-large, every whitening level trails raw PQ while the zero-cost OPQ rotation wins outright).
\item \textbf{Where the correction can be learned (Sects.~\ref{sec:learning} and \ref{sec:evaluation}).} Controlled probes on a frozen encoder answer the learning question in both directions. Geometry corrections belong post-hoc: no learned variant beats the specialized solvers, and each probe's outcome lands on the map's own curves. The one correction that belongs in the encoder is score margins --- a statistic no post-hoc transform can sharpen: a single trainable final layer nearly triples them, lifting every index family's recall at fixed budget for a measured 3--5-point task price. For the navigation families, which serve changing corpora and have almost no post-hoc lever, this is the one demonstrated lever --- it ships with the model, nothing to refit per corpus; training it jointly with the task objective is the open next step (Sect.~\ref{sec:discussion}).
\end{enumerate}

\section{Related Work}\label{sec:related}

\paragraph{Embedding anisotropy.} The cone effect and representation degeneration
\cite{gao2019degeneration,ethayarajh2019contextual}; post-hoc fixes --- all-but-the-top
\cite{mu2018allbut}, BERT-flow \cite{li2020bertflow}, whitening \cite{su2021whitening};
rogue dimensions \cite{timkey2021rogue}; and the counterpoint that
isotropy needn't help downstream tasks (I-STAR~\cite{rudman2024istar}) ---
whose companion critique, that average-cosine measures are flawed \emph{as
isotropy scores}, does not touch this paper's use of MCS: it enters not
as an isotropy score but as the mean-direction statistic that Lemma~\ref{lem:contraction} and
Theorem~\ref{thm:degeneration} tie mechanistically to sketch collisions, while spectrum-level
questions are carried by spectral flatness and the participation ratio.
This literature measures exact-scoring and downstream quality. Read
through Sect.~\ref{sec:axes}'s decomposition, it documents the semantic-shift component:
the cone component itself cannot move an exact ranking (Lemma~\ref{lem:invariance}), so
what these interventions move, they move by rewriting the encoder's
semantics --- which is why the reported sign varies by encoder, task, and
reference frame (Jung et al.\ \cite{jung2023isotropic} report whitening \emph{helping} dense
retrieval; measured here, the same transform \emph{rewrites} 62\% of true
neighbors; both are correct, in different frames, Sect.~\ref{sec:groundtruth}). What this line
does not measure is the approximation layer, where the same geometry
acts lawfully and per-family (Sect.~\ref{sec:families}); predicting the two components
together is this paper's subject.

\paragraph{Indexing and quantization.} PQ \cite{jegou2011pq} and OPQ
\cite{ge2013opq}, whose Gaussian analysis the misalignment factor of Sect.~\ref{sec:pq} repackages into a
measurable, factored statistic; ScaNN's anisotropic quantization loss
\cite{guo2020scann}, orthogonal to \emph{data} anisotropy; RaBitQ \cite{gao2024rabitq},
production evidence that fixed-grid corrections ship as rotations; MUVERA
\cite{dhulipala2024muvera}, whose data-oblivious FDE guarantee Theorem~\ref{thm:degeneration} analyzes
(Theorem~\ref{thm:degeneration}); IVF and centroid-residual systems
(ColBERTv2/PLAID~\cite{santhanam2022colbertv2,santhanam2022plaid}); HNSW \cite{malkov2018hnsw}.

\paragraph{Hardness and local structure.} Hubness \cite{radovanovic2010hubs} and its
centering-based reductions for cosine pipelines \cite{suzuki2013centering,hara2015localized};
local intrinsic dimensionality and its role in nearest-neighbor
hardness and benchmark difficulty \cite{houle2013dimensionality,amsaleg2015lid,aumuller2021local}.
On the theory side, guarantees for
graph-based search exist only in restricted regimes: Prokhorenkova \&
Shekhovtsov \cite{prokhorenkova2020graph} analyze the low-dimensional regime ($d \ll \log n$),
and Indyk \& Xu \cite{indyk2023worstcase} prove guarantees under bounded intrinsic dimension
for one DiskANN variant while exhibiting hard instances on which HNSW and
NSG need near-linear query time. Lakshman et al.\ \cite{lakshman2025stability} develop a
stability theory for modern vector retrieval and prove, concurrently
with this paper's Theorem~\ref{thm:degeneration} and by a different route, that Chamfer/MaxSim
similarity preserves stability while average-pooling aggregation may
destroy it --- the stability-theoretic form of Sect.~\ref{sec:fde}'s collision-collapse
mechanism, without the anisotropy trigger or measured calibration
supplied here. This line is the closest prior to the
prediction question; it concentrates on graph/partition methods,
observational difficulty measures, and worst-case constructions. This
paper differs in covering fixed-grid
quantization under the same roof, in validating predictions
out-of-sample rather than by in-sample correlation, and in the twin
hierarchy (Sect.~\ref{sec:twin}), which turns ``which structure matters'' into a controlled
comparison instead of a regression.

\paragraph{Index-aware training and index-side learning.} Encoder--index
co-optimization is established art: Poeem and JPQ train retrieval
representations jointly with PQ-based indexes \cite{zhang2021poeem,zhan2021jpq},
JTR co-trains a tree index with the query encoder \cite{li2023jtr},
and EHI learns a hierarchical index with a dual encoder end to end
\cite{kumar2024ehi}; CRISP trains inherently clusterable token
representations directly into a late-interaction encoder \cite{veneroso2025crisp}
--- training-time shaping of exactly the cluster structure this
paper's navigation analysis measures. These systems put a \emph{specific index inside
a supervised training loop}, coupling the encoder to that index and
corpus.
The learning result is complementary and different in kind: it identifies a
\emph{transferable statistic} --- score margins --- that lifts every index family
at once, is movable label-free with no index in the loop, and is exactly
the kind of quantity this co-training machinery could adopt as a target.
On the index side, optimistic routing estimates shard maxima from
within-shard moments and LIRA learns query-aware probing
\cite{bruch2025optimistic,zeng2025lira} --- improvements to navigation indexes
from inside the index, consistent with this paper's measurement that the
embedding-space side offers these families almost no post-hoc lever.
These are per-query, index-in-hand decisions, not pre-build forecasts;
and optimistic routing's effective statistics are \emph{shard-conditional}
moments --- cluster-resolved information, the same resolution level the
cluster panel and twin preserve and global moments provably lack
(Sect.~\ref{sec:measurement}).
Elliott \& Clark \cite{elliott2024hnsw} show HNSW recall shifts by up to 12 points with
insertion order and local intrinsic dimensionality --- independent evidence
for Sect.~\ref{sec:hnsw}'s account that graph behavior rides on structure summary
statistics do not capture.

\paragraph{Positioning.} No known prior work: (i) predicts index-layer
behavior across fixed-grid, partition, and graph families from one
label-free measurement pass; (ii) validates those predictions with
out-of-sample forecasts on held-out encoder--dataset pairs; (iii) separates every corrective
transform's effect into fidelity spent and efficiency gained; or (iv) uses
structure-matched synthetic twins both to attribute index behavior to
specific layers of geometric structure and as zero-parameter simulators
that predict absolute recall.

\section{The Prediction Framework: Instruments and the Assembled Ladder}
\label{sec:framework}

This section builds the predictive ladder in parts, then assembles it.
Sects.~\ref{sec:groundtruth}--\ref{sec:labels} present the components: the ground truth every prediction is
scored against (Sect.~\ref{sec:groundtruth}), the two axes every transform trades between
(Sect.~\ref{sec:axes}), the statistics that do the predicting and the closed forms
they feed (Sect.~\ref{sec:measurement}), the twin
instrument that carries prediction where formulas fail (Sect.~\ref{sec:twin}), and the
labels that grade every claim's strength (Sect.~\ref{sec:labels}). Sect.~\ref{sec:setup} fixes the
corpora, encoders, and index configurations used throughout, and Sect.~\ref{sec:ladder}
assembles the components into the full ladder --- the operating procedure
the rest of the paper validates.

\subsection{Ground truth: the encoder's contract}
\label{sec:groundtruth}

Every measurement in this paper is anchored to one fixed ground truth:
the \emph{original} embedding space's exact top-$k$ --- the contract the encoder
was evaluated under. This choice is what makes the framework deployable
and falsifiable. It is label-free, so any corpus--encoder pair can be
measured. It is complete for indexes, which never move points and so can
only fail to \emph{recover} this truth (Sect.~\ref{sec:axes}'s efficiency). And it is complete
for corrective transforms, whose entire effect on the contract is
captured by how much of the truth survives them (Sect.~\ref{sec:axes}'s fidelity) plus
how easily the index recovers what remains. A task-blind transform should not be expected to make the rewritten
ranking \emph{semantically} better: anything a fixed corpus-level map could
add to task relevance, the encoder's own training could have absorbed.
Where such transforms have helped task metrics in the literature (Sect.~\ref{sec:related}),
it marks a mismatch between how the encoder was trained and how it is
used --- a repair of a training deficiency, not a property to plan on. The
paper therefore holds the encoder's contract fixed and prices every
transform against it.

This label-free ground truth suffices for indexes and for blind
transforms. It stops sufficing at the one point where the encoder side
itself is modified: the learned transforms of Sect.~\ref{sec:learning} are \emph{trained} to
preserve the measurement's own ground truth --- the encoder's exact
top-$k$ --- so measuring them against it cannot reveal what the training
traded away --- by construction they score near-perfectly. There, and
only there, the transformed encoder's \emph{exact} retrieval is additionally
checked against human relevance judgments (qrels) --- a second
reference frame called the \textbf{task frame}, marked wherever it
appears --- as a regression check that the new exact top-$k$ still
retrieves what is truly relevant: that optimizing the label-free proxy
has not drifted the semantics the proxy stands for. The approximation
layer plays no role in this check; it concerns the transformed
encoder's exact output alone. The check earns its keep: one Sect.~\ref{sec:learning} probe
scores near-perfectly in the very frame its training targeted,
while at exact scoring the task frame registers a six-point loss of
top-rank relevance (Sect.~\ref{sec:learning} iii) that the
label-free frame, by construction, cannot see (Sect.~\ref{sec:learning}).

\subsection{Fidelity, efficiency, and the exchange bound}
\label{sec:axes}

\begin{figure}[t]
\includegraphics[width=\textwidth]{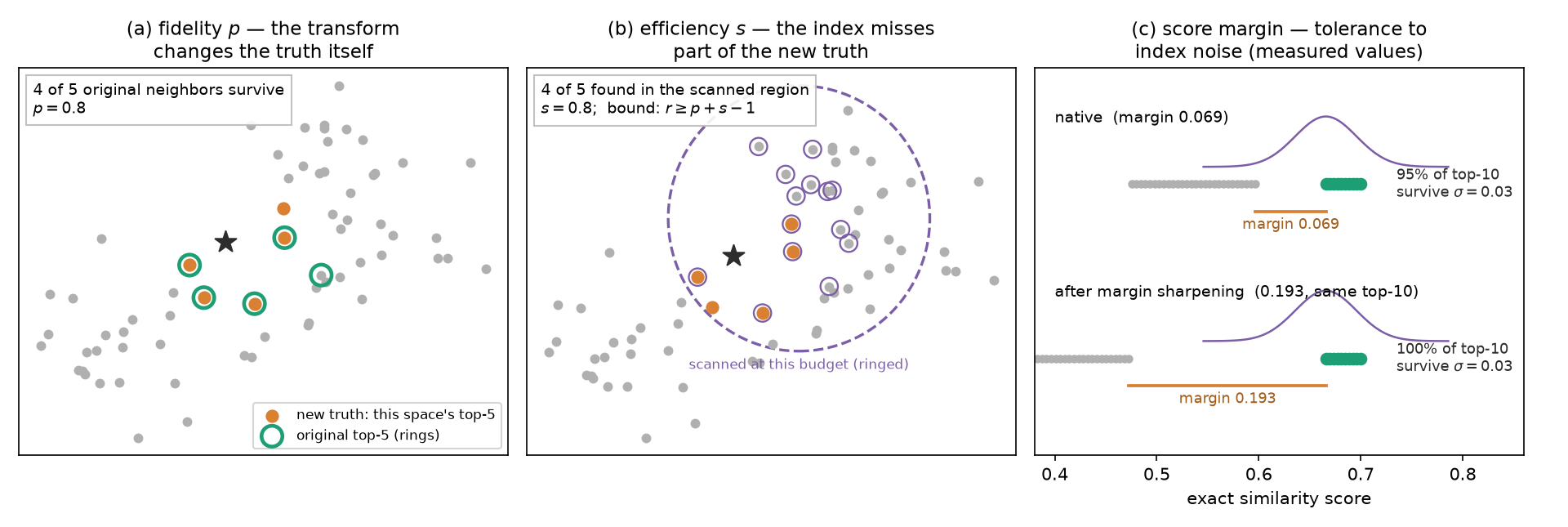}
\caption{The three core instruments, schematically. \emph{(a)} A corrective
transform moves points, so the transformed space's own exact top-$k$ (filled)
is no longer the original top-$k$ (rings): the overlap is fidelity $p$.
\emph{(b)} Within the transformed space, an index at a fixed budget scans only
part of the corpus (ringed): the fraction of the space's own top-$N$ it
finds is efficiency $s$; end-to-end recall obeys $r \ge p + s - 1$ at
matched depth ($k = N$). \emph{(c)} The score margin --- the gap between the
$k$-th exact score and the deeper field --- is the ranking's tolerance to
the score noise an index injects; the two strips use this paper's
measured margins (native $0.069$ vs margin-sharpened $0.193$, same
top-10), and survival rises with the margin-to-noise ratio (the Sect.~\ref{sec:fidelity}
gap model).}
\label{fig:framework}
\end{figure}

With the ground truth fixed, the two axes it induces can be defined
exactly (Fig.~\ref{fig:framework}a--b; panel c previews the margin
instrument of Sect.~\ref{sec:measurement}). For a corrective transform
$T$ applied to corpus and queries, with base
exact top-$k$ set $B$, the transformed space's own exact top-$N$ set $K_N$,
and the index's returned candidate set $C$ at a fixed compute budget
($|C| = N$):

\begin{itemize}
\item \textbf{Fidelity} $p_N = |B \cap K_N| / |B|$ --- how much of the original truth
  survives the transform. An orthonormal rotation preserves every inner
  product and hence every exact ranking, so isometries have $p_N \equiv 1$
  by construction; what each distorting transform spends is measured in
  Sect.~\ref{sec:calculus}.
\item \textbf{Efficiency} $s_N = |K_N \cap C| / N$ --- how much of the transformed
  space's \emph{own} truth the index returns; pure index quality, uncontaminated
  by the geometry change.
\item \textbf{End-to-end recall} $r = |B \cap C| / |B|$ --- coverage of the
  original truth by what the index returns. The two axes bound it: exactly
  $p_N k$ members of $B$ lie in $K_N$, and the index misses at most
  $N(1 - s_N)$ members of $K_N$, so $|B \cap C| \ge p_N k - N(1 - s_N)$;
  dividing by $|B| = k$ gives $r \ge p_N - (N/k)(1 - s_N)$, and at matched
  depth ($k = N$) $r \ge p_N + s_N - 1$. The bound is set-theoretic and
  exact, and tight exactly where the derivation forces equality ---
  untransformed spaces, where $p = 1$; Sect.~\ref{sec:protocol} verifies
  it is never violated across this paper's measurements.
\end{itemize}

A correction helps only if the efficiency it buys outweighs the fidelity it
spends --- the exchange this paper quantifies and predicts.

\subsection{The master measurement}
\label{sec:measurement}

Fidelity and efficiency are what the framework predicts; what it
predicts \emph{from} is one label-free pass over an embedding sample, which
records two kinds of statistics.

\textbf{The analytic core: mean $\mu$ and covariance $\Sigma$}, read at four
resolutions, each with a designated consumer:

\begin{center}
\footnotesize
\setlength{\tabcolsep}{4pt}
\begin{tabular}{@{}p{0.25\textwidth}p{0.30\textwidth}p{0.35\textwidth}@{}}
\toprule
resolution of $(\mu, \Sigma)$ & statistic & consumer \\
\midrule
eigenvalues (basis-free) & spectral flatness, participation ratio & the rotation-invariant floor: best-achievable fixed-grid behavior \\
blocks in the operating basis & subvector determinants, cross-block correlation $\to$ \textbf{misalignment factor} (Sect.~\ref{sec:pq}) & PQ \\
diagonal in the operating basis & variance balance, kurtosis & scalar/binary quantization (recorded; not analyzed in this paper) \\
$\mu$ against $\Sigma$'s scale & mean cosine similarity (MCS) & FDE/SimHash (Sect.~\ref{sec:fde}) \\
\bottomrule
\end{tabular}
\end{center}

(Spectral flatness is the geometric-to-arithmetic mean ratio of
$\Sigma$'s eigenvalues; the participation ratio, $(\sum_i \lambda_i)^2 /
\sum_i \lambda_i^2$, is an effective dimension count.)

Each analytic-core row feeds a \emph{closed form} --- a formula, derived in
Sect.~\ref{sec:families}, that maps the statistic directly to a family's behavior: Theorem~\ref{thm:degeneration}
for the FDE grid and the misalignment factor for PQ (IVF's routing
admits a derived form with fitted constants, Sect.~\ref{sec:ivf}). Closed forms are
the ladder's first instrument class: evaluated at a measured seed they
give the verdict, and evaluated along their statistic they extrapolate
it to geometry not yet encountered (Fig.~\ref{fig:counterfactual}, left). The second kind of
statistic requires the second instrument class.

\textbf{Measured statistics that moments provably do not determine} (established
by the twin comparison, Sect.~\ref{sec:twin}): the score-margin distribution (gaps between
the $k$-th and deeper exact neighbors --- how much perturbation a ranking
tolerates); the cluster panel --- intra-cluster distance, inter-centroid
distance, their ratio, the \textbf{routing margin} (each point's gap between its
nearest and second-nearest cluster centroid --- the direct measure of how
unambiguously each point routes), and their twin-normalized ``excess''
versions; and the local-structure panel (measured two-NN intrinsic
dimension, hubness statistics, local contrast, distance concentration).

These instruments have close relatives in prior literature, and the
names are chosen deliberately. The score margin is the per-ranking analogue of
\emph{relative contrast} \cite{he2012difficulty}, the classic hardness measure
for nearest-neighbor search, and of the distance-concentration quantities
underlying local intrinsic dimensionality \cite{houle2013dimensionality}. The routing margin
is a centroid-based, per-point variant of the \emph{silhouette} width
\cite{rousseeuw1987silhouettes}: where the silhouette compares mean distances to whole
clusters, the routing margin compares distances to centroids --- because a centroid
comparison is literally the operation IVF routing performs, so the
statistic measures exactly the decision the index will make. The
inter/intra ratio and Cali\'nski--Harabasz \cite{calinski1974dendrite} index are standard
cluster-validity measures, recorded unmodified. Table~\ref{tab:instruments} collects the
paper's instruments, their nearest relatives, and their consumers.

\begin{table}[t]
\caption{The instrument panel.}
\label{tab:instruments}
\footnotesize
\setlength{\tabcolsep}{4pt}
\begin{tabular}{@{}p{0.16\textwidth}p{0.30\textwidth}p{0.24\textwidth}p{0.22\textwidth}@{}}
\toprule
instrument & definition & nearest relative & consumed by \\
\midrule
fidelity $p_N$ & original exact top-$k$ surviving in the transform's own top-$N$ & --- (this paper's frame split) & every transform verdict (Sect.~\ref{sec:calculus}) \\
efficiency $s_N$ & transform's own top-$N$ found by the index at budget & ann-benchmarks recall, per-frame & every index curve (Sect.~\ref{sec:families}) \\
score margin & gap from $k$-th to deeper exact scores & relative contrast \cite{he2012difficulty} & fidelity model (Sect.~\ref{sec:fidelity}); HNSW (Sect.~\ref{sec:hnsw}) \\
MCS & mean cosine similarity (cone strength) & anisotropy \cite{ethayarajh2019contextual} & FDE/SimHash (Sect.~\ref{sec:fde}) \\
misalignment factor & $\mathrm{MF} = \frac{1}{M}\sum_j \lvert\Sigma_j\rvert^{1/d_s} / \lvert\Sigma\rvert^{1/d}$ & --- (derived here, Appendix~\ref{app:proofs}) & PQ (Sect.~\ref{sec:pq}) \\
routing margin & centroid gap $(d_2 - d_1)/d_1$ per point & silhouette \cite{rousseeuw1987silhouettes}, centroid variant & IVF (Sect.~\ref{sec:ivf}) \\
inter/intra ratio, CH & cluster separation vs spread & cluster validity \cite{calinski1974dendrite} & cluster panel, twin tolerance (Sect.~\ref{sec:twin}) \\
cluster excess & Cali\'nski--Harabasz index divided by its value on the moment-matched twin & --- (twin-normalized here) & IVF refit (Sect.~\ref{sec:round1}) \\
\bottomrule
\end{tabular}
\end{table}

\subsection{The twin hierarchy}
\label{sec:twin}

The split in Sect.~\ref{sec:measurement} --- statistics that admit formulas versus statistics
the moments provably do not determine --- is established by the
instrument defined here. To decide which layer of structure an index
consumes, each real corpus is compared against synthetic twins that
match successively more structure --- exactly by construction at the
first tier, tolerance-checked at the second:

\smallskip
\noindent\textbf{Tier 1 --- moment-matched twin}: sampled from a Gaussian with the corpus's
$(\mu, \Sigma)$. Identical first and second moments by construction; any
measured difference is structure beyond moments.

\smallskip
\noindent\textbf{Tier 2 --- cluster-matched twin}: real cluster weights, centers, and
within-cluster covariances; Gaussian \emph{inside} each cluster. Because
this tier's match is estimated rather than exact, it carries a
matching tolerance: the cluster panel of Sect.~\ref{sec:measurement}, measured on the sampled
twin itself, must reproduce the real corpus's, each statistic within
2\% (met in every experiment at the $25$--$50$k scales studied; the one
exception is the full-size million-document twins of
Sect.~\ref{sec:milliondoc}, where three of four encoders exceed it --- a
failure that is itself diagnostic). With
the tolerance met, any measured real-versus-twin difference in index
behavior is attributable to structure beyond cluster-level
organization --- the license the attribution below depends on.

\begin{figure}[t]
\includegraphics[width=\textwidth]{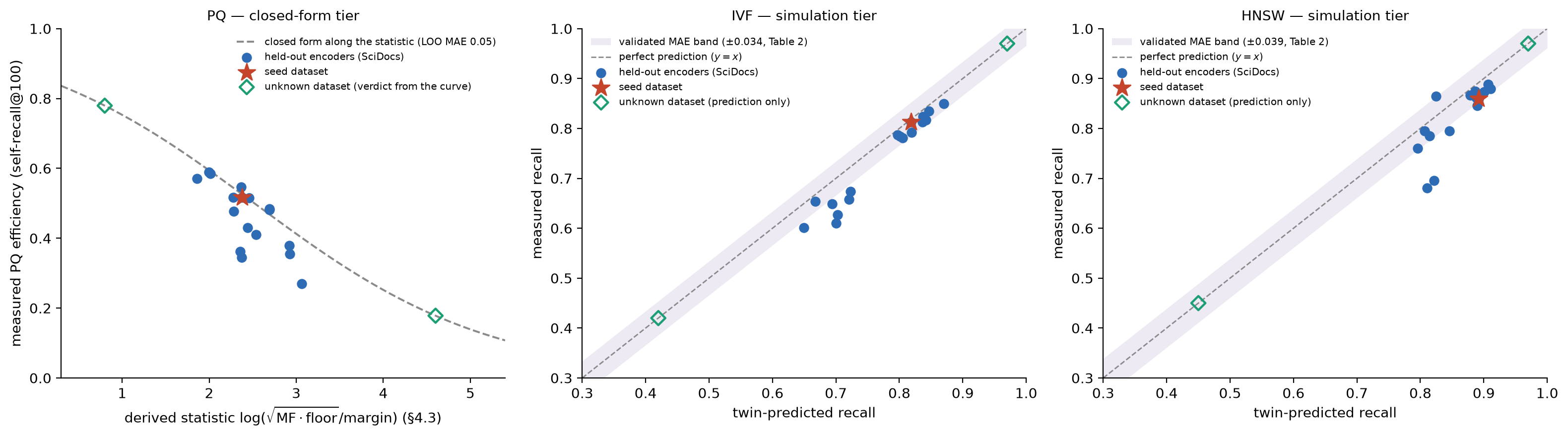}
\caption{One seed --- verdicts for datasets that do not exist yet,
across all three measured families. The same seed dataset (starred)
anchors every panel. \emph{Left}, the closed-form tier: PQ efficiency
against the derived statistic of Sect.~\ref{sec:pq}; the dashed curve is the
validated closed form itself (leave-one-encoder-out MAE $0.05$,
Sect.~\ref{sec:round1}), so an
unknown dataset (open diamonds) reads its verdict off the curve at its
own measured statistic --- including beyond the range any measured pair
occupies. \emph{Middle and right}, the simulation tier: absolute navigation
levels admit no closed form (Sect.~\ref{sec:hnsw}), so the verdict comes from minting
the twin --- twin-predicted against measured recall for the $17$
held-out encoders on SciDocs (document-proxy configuration), with
the validated MAE of Sect.~\ref{sec:twin25k} (Table~\ref{tab:fullcorpus}) as the band; an unknown dataset arrives with
only its predicted coordinate. IVF's low-recall cluster sits below the
band --- the $+0.03$ over-prediction bias documented in Sect.~\ref{sec:twin25k} (Table~\ref{tab:fullcorpus}), visible.}
\label{fig:counterfactual}
\end{figure}

\smallskip
The hierarchy serves twice. As an \emph{attribution instrument}, the tier-to-tier
gaps decompose an index's behavior into contributions of moments, cluster
structure, and a residual. As a \emph{predictor}, a tier-2 twin is a
zero-fitted-parameter simulator with three operating modes: full-size
minting (hosts the target's true index geometry; also the attribution
instrument); the size ladder (multi-draw, multi-size extrapolation ---
the recommended default for level prediction at large corpus scale,
Sect.~\ref{sec:milliondoc}); and \emph{counterfactual minting} --- hold a seed sample's fitted
statistics and dial them (cluster separation, margins, the mean
direction), re-minting at each setting, so that a single reference
sample maps how every index family responds to geometry it has not yet
encountered. The third mode serves the common case where the production
corpus does not exist yet, or is expected to drift from the sample in
hand; the closed-form families support the same exploration directly,
by evaluating their formulas along the statistic --- Figs.~\ref{fig:fde}--\ref{fig:ivf} are
exactly such response curves. Fig.~\ref{fig:counterfactual} demonstrates the mode on all
three measured families: from one seed dataset, verdicts issue for
datasets that do not yet exist --- off the closed form's curve for PQ,
from minted twins for IVF and HNSW. The causal sweeps of Sect.~\ref{sec:fde} and the
single-statistic probes of Sect.~\ref{sec:learning} validate the mode: geometry dialed while
semantics are held fixed, with each family's recall responding as its
governing statistic moves. Query source is an explicit axis of the
instrument: twin
queries default to the document mixture because a query sample of a few
hundred cannot fund its own statistics --- the scale asymmetry between the
two sides of retrieval --- and Sect.~\ref{sec:twin25k} measures exactly what this proxy costs
and when a real query log should replace it.

\subsection{Claim labels}
\label{sec:labels}

Instruments this different --- proofs, closed forms, simulators --- yield
claims of different strength, so every predictive claim carries one of
three labels. \textbf{proved} --- exact
within the stated model; tested by invariant checks (a violation indicates an implementation
error). \textbf{derived} --- a quantitative prediction under listed assumptions,
closed-form or run-on-a-twin; tested by calibration on held-out pairs; demoted if residuals show
structure.
\textbf{measured} --- a reproducible regularity without a surviving derivation;
still required to issue falsifiable forecasts. Nothing is labeled by
aspiration; several demotions and one promotion occur in Sect.~\ref{sec:evaluation}.

\subsection{Experimental setup}
\label{sec:setup}

Every measurement in the paper runs on one fixed bed. Corpora: MS MARCO passage shards ($50$k in-sample; $25$k held-out with
qrels) and BEIR SciFact, NFCorpus, SciDocs, ArguAna; $17$ dense and $4$
late-interaction encoders (roster in Appendix~\ref{app:details}); indexes: faiss PQ
($M{=}16$, 8-bit), OPQ, IVF (nlist $1024$, nprobe $16$ unless a grid is
stated), HNSW ($M{=}32$, efConstruction $100$, efSearch grids),
spec-faithful MUVERA FDE; depths $k{=}10$, $N{=}100$; scale tests at
$1{,}001{,}595$ HotpotQA documents. Full configurations: Appendix~\ref{app:details}.

\subsection{The ladder, assembled}
\label{sec:ladder}

The instruments compose into a single operating procedure --- the paper's
predictive ladder --- run entirely before any production index is built.
Each step names the section that licenses it; the two empirically set
choices, the family routing and the mode-selection rule, are validated
in Sects.~\ref{sec:families} and \ref{sec:evaluation}.

\begin{enumerate}
\item \emph{Measure.} Run the label-free pass of Sect.~\ref{sec:measurement} on the raw embeddings:
   the moment resolutions, score margins, and the cluster panel.
\item \emph{Read the fixed-grid verdicts from closed forms.} MCS orders FDE
   behavior and decides centering's sign (Sect.~\ref{sec:fde}); the misalignment
   factor decides whether a rotation pays and prices PQ (Sect.~\ref{sec:pq}).
\item \emph{Price any candidate transform.} Predict fidelity from the gap model
   (Sect.~\ref{sec:fidelity}) and efficiency from the post-transform statistics; the
   exchange bound (Sect.~\ref{sec:axes}) composes the two into an end-to-end verdict
   per index.
\item \emph{For partition and graph indexes, mint the twin} (Sect.~\ref{sec:twin}): fit the
   cluster mixture, verify the matching tolerance, generate. Use a real
   query log where one exists, the document proxy otherwise (Sect.~\ref{sec:twin25k}
   prices the difference).
\item \emph{Choose the operating mode by family and scale.} For IVF, mint at
   full size at every scale validated in Sects.~\ref{sec:twin25k}--\ref{sec:milliondoc} --- the full-size twin
   hosts the target's true cell structure. For HNSW, mint at full size
   up to the $\sim 50$k scales validated there; beyond that --- or
   whenever the matching tolerance cannot be met at the target size ---
   mint at sizes where it is met, extrapolate the measured size trend,
   and calibrate against one measured encoder (a
   $\sim 25{,}600$-document statistics budget sufficed at one million
   documents; Sect.~\ref{sec:milliondoc}).
\item \emph{Simulate.} Build the intended index, at its intended parameters
   (nlist, nprobe, $M$, efSearch), on the twin and measure recall: that
   number is the forecast, with expected error given by Tables~\ref{tab:fullcorpus}--\ref{tab:million} and
   each claim carrying its Sect.~\ref{sec:labels} label.
\item \emph{Explore counterfactuals (optional).} When the production corpus is
   unavailable or expected to drift from the sample in hand, the
   measured sample of step 1 is the \emph{seed} (Sect.~\ref{sec:twin}): hold it fixed and
   dial the governing statistics --- re-minting twins at perturbed
   cluster parameters, or evaluating the closed forms along their
   statistic --- to map each family's compatibility across the geometry
   range production may occupy. Fig.~\ref{fig:counterfactual} shows both forms: verdicts
   read off the closed form's curve, and twin predictions issued for
   datasets far from the seed, before any measurement exists.
\end{enumerate}

\section{Per-Family Analysis}\label{sec:families}

This section answers the prediction question --- which measurable properties
determine each index family's behavior --- one family at a time, each claim
carrying its Sect.~\ref{sec:labels} label.

\subsection{Exact scoring is invariant (proved)}\label{sec:exact}

In the $\rho$-cone model --- every embedding a shared unit mean direction
of weight $\rho$ plus an orthogonal isotropic residual of weight
$\sqrt{1-\rho^2}$, so embeddings are unit-norm and MCS $= \rho^2$
(Appendix~\ref{app:proofs}) --- exact inner-product/MaxSim ranking is
independent of the common-mean component (Lemma~\ref{lem:invariance}):
exact benchmarks cannot see the phenomena this
paper studies. The invariance requires no re-normalization; real pipelines
that center and re-normalize \emph{do} perturb rankings --- measurably
(Sect.~\ref{sec:calculus}), which is the fidelity cost.

\subsection{FDE / SimHash (derived + theorem; measured on real encoders)}\label{sec:fde}

Lemma~\ref{lem:contraction}: the common-mean cone contracts every semantic
angle by $\sqrt{1-\mathrm{MCS}}$ (in the $\sin(\theta/2)$ sense) before
SimHash sees it. Theorem~\ref{thm:degeneration}
(Appendix~\ref{app:proofs}): at any fixed FDE budget, sufficient anisotropy
forces full bucket collision, collapsing the encoding to mean-pooling and
provably inverting rankings exact MaxSim gets right; and at fixed FDE
parameters, the additive error MUVERA's data-oblivious guarantee delivers
on the ranking-relevant residual scores carries a hidden
$(1-\mathrm{MCS})^{-1}$ (Corollary~\ref{cor:inflation}).

Whether collision \emph{hurts} depends on whether relevance is recoverable
from document means. Two synthetic token corpora --- MCS dialed while
semantics are held fixed, relevance planted by construction
(Appendix~\ref{app:details}) --- separate the two regimes. On the
coarse-relevance corpus (relevance =
topic-set overlap), spec-faithful FDE barely degrades (recall $0.989 \to
0.922$ as MCS $\to 0.91$) even as collisions reach $0.92$. On a
fine-grained-relevance corpus (a minority of tokens carries the signal ---
the realistic retrieval regime), the causal tax appears: $0.897 \to 0.763$,
centering flat at $0.885$, exact ranking bit-invariant ($5$ seeds). On real
GTE-ModernColBERT tokens (MCS $0.910$, $25{,}000$ passages), the tax is
large: raw recall@100 $= 0.136$, recovered $3.4\times$ by centering
($0.468$) at fidelity cost $p = 0.924$; whitening adds nothing over
centering ($s = 0.555$ for both) and costs more ($p = 0.849$). A
pooling-implementation control on identical data shows the legacy
max-by-norm aggregation would have overstated the raw tax ($0.054$),
motivating the spec-faithful measurements throughout.

\begin{figure}[t]
\includegraphics[width=\textwidth]{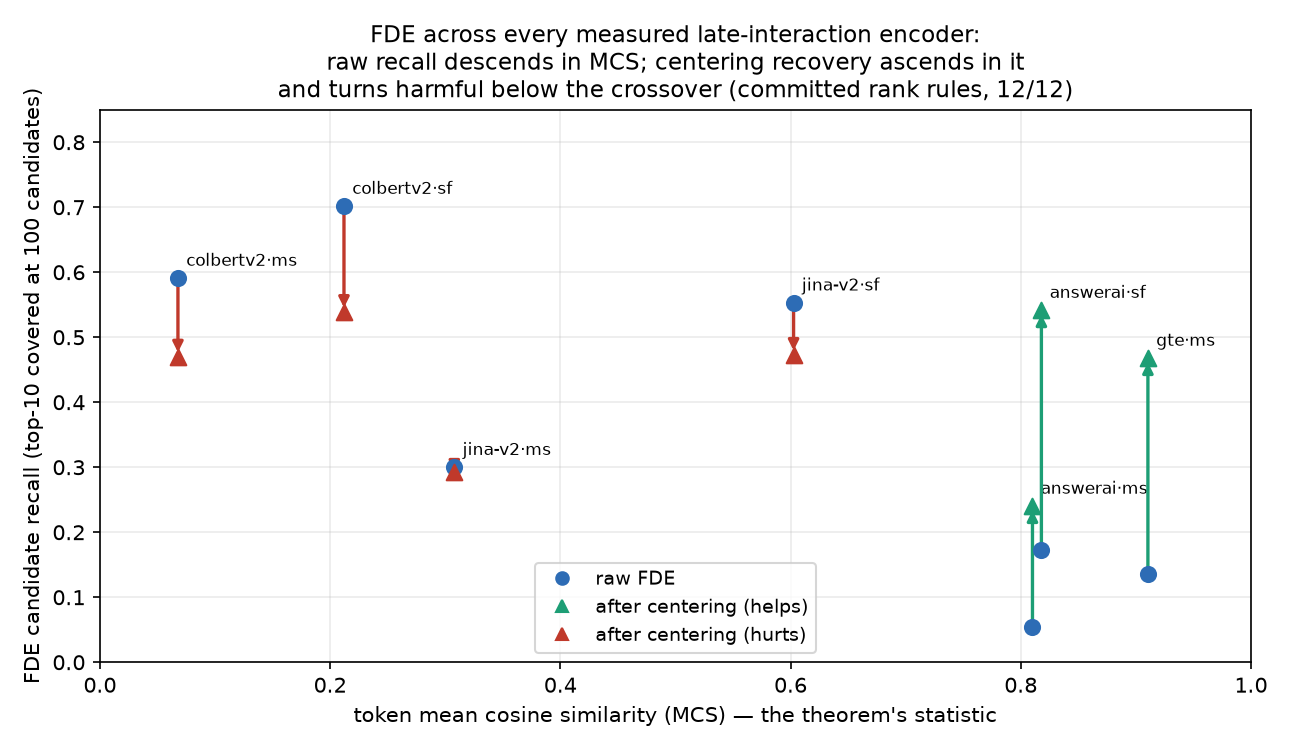}
\caption{FDE behavior across all seven measured encoder--dataset pairs.
Each blue point is raw FDE candidate recall --- coverage of the exact
top-$10$ by the $N{=}100$ returned candidates, Sect.~\ref{sec:axes}'s
end-to-end $r$ --- at the encoder's token MCS;
the arrow shows the effect of centering (green up, red down). The two
rank rules --- raw recall descends in MCS, centering recovery
ascends in it --- are the visible pattern, and centering flips harmful for
the low-MCS encoders on the left.}
\label{fig:fde}
\end{figure}

Across every measured late-interaction encoder, the statistic orders
behavior exactly as the theorem directs (Fig.~\ref{fig:fde}): raw FDE
recall descends in token MCS and centering recovery ascends in it --- both
fixed as rank rules during development and confirmed on all $12$
ordered held-out pairs (Sect.~\ref{sec:fderules}) --- including the
decision-relevant sign result that centering \emph{harms} the low-MCS
encoders (ColBERTv2: $-0.12$ / $-0.16$). The correction every deployment
guide recommends is not merely wasted on an already-isotropic encoder; it
is harmful, and one measured statistic predicts which side of the line an
encoder falls on. The tax persists at scale: on a nested
$100{,}593$-document sample, GTE-ModernColBERT's raw versus centered
contrast holds ($0.009$ vs $0.093$), and the centering sign is stable
under doubled hashes, repetitions, and projection width.

\begin{figure}[t]
\includegraphics[width=\textwidth]{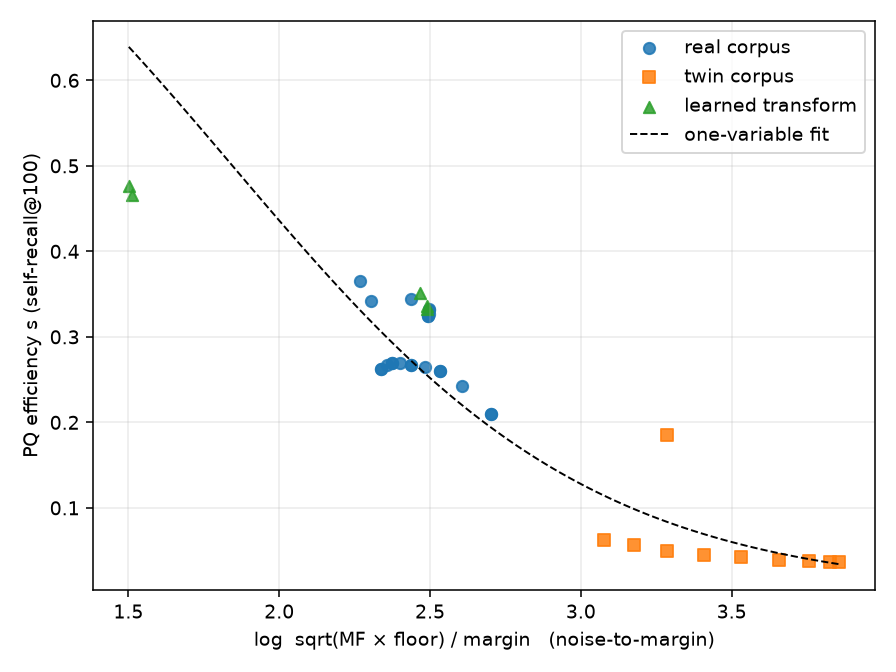}
\caption{PQ closes in closed form. Efficiency across 37 real, twin,
and learned-transform states versus the derived statistic
($R^2=0.87$; a transform-family indicator adds only $+0.042$ --- the
statistic, not the transform's identity, carries the prediction).}
\label{fig:pq}
\end{figure}

\subsection{PQ: the misalignment factor (derived)}\label{sec:pq}

For $M$ subvectors of dimension $d_s$ under equal per-block rate --- equal
rate is structural in PQ, and is exactly why imbalance costs --- high-rate
Gaussian quantization~\cite{zador1982asymptotic,gersho1979asymptotically}
gives total distortion $\propto \sum_j |\Sigma_j|^{1/d_s}$. Two classical inequalities bound its rotation
floor --- the rotation-invariant minimum $M|\Sigma|^{1/d}$ of the
distortion sum (Appendix~\ref{app:proofs}):
AM--GM (equality iff block determinants balance) and Fischer's inequality
(equality iff no cross-block correlation). Hence the \textbf{misalignment
factor}

\[
\mathrm{MF} \;=\; \frac{\tfrac{1}{M}\sum_j |\Sigma_j|^{1/d_s}}{|\Sigma|^{1/d}}
\;=\; [\text{AM--GM gap}]\times[\text{Fischer gap}] \;\ge\; 1,
\]

computable from the empirical covariance alone, with OPQ~\cite{ge2013opq} as basis
optimization over the same object (its practical solver lands near, not at,
the floor). Diagnosis on BGE-large/MS MARCO: $\mathrm{MF} = 4.43 = 1.00
\times 4.43$ --- the real PQ tax is \emph{pure cross-subvector correlation},
zero variance imbalance, which is why trace-based balance statistics read
``fine'' while OPQ gains $42\%$ ($0.575 \to 0.817$ candidate recall at
fidelity $1.0$). Because a rotation preserves every inner product, it
leaves all score margins unchanged --- so OPQ is a margin-controlled
causal check: applied in the same space it cuts MF to $1.80$ and PQ
efficiency rises $0.324 \to 0.481$; the statistic moves, the margin
cannot, and efficiency follows.
Efficiency follows $\mathrm{logit}(s) \approx \beta_0 + \beta_1
\log(\sqrt{\mathrm{MF}\cdot\mathrm{floor}}/\mathrm{margin})$ (margin:
the score margin of Sect.~\ref{sec:measurement}) --- one curve
(Fig.~\ref{fig:pq}) fits $37$ states (a state: one corpus--transform configuration) spanning
real, twin, and learned transforms ($R^2 = 0.87$, the fraction of
variance in $\mathrm{logit}(s)$ the fit explains; a transform-family
indicator adds $+0.042$).

\subsection{IVF: routing as gap survival (derived form; constants fitted)}\label{sec:ivf}

\begin{figure}[t]
\includegraphics[width=\textwidth]{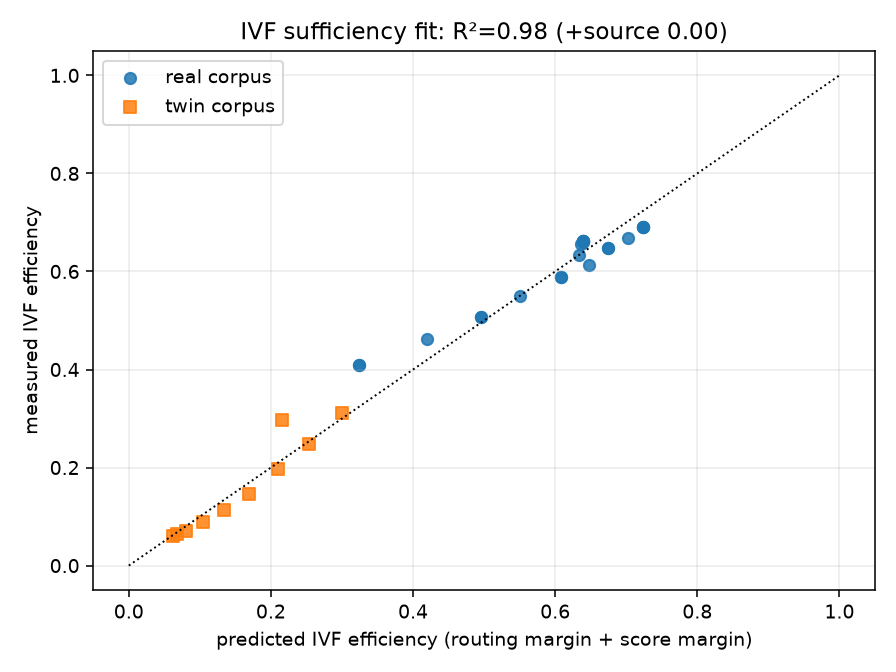}
\caption{IVF follows its derived form. Measured efficiency against
the gap-survival prediction (routing margin + score margin) across 32
real and twin states, $R^2 = 0.98$ in-family; both corpus types land on
one line. The form supplies the lever and the ordering; absolute levels
for unseen encoders come from twin simulation (Fig.~\ref{fig:navclosure},
Sect.~\ref{sec:twin25k}).}
\label{fig:ivf}
\end{figure}

IVF recall factors into routing success --- whether the document's cell
is among the probed cells --- and in-cell ranking. Modeling a query as its neighbor plus a perturbation,
routing success is a gap-survival problem at the centroid level --- the same
structure as the fidelity model of Sect.~\ref{sec:fidelity}, one level up ---
giving $\mathrm{logit}(s) \approx c_0 + c_1\log(\text{routing margin}) +
c_2\log(\text{score margin}) + c_3\log(\text{cluster excess})$, each
partial effect positive by mechanism. The marginal data agree (routing
margin alone: $\rho = 0.93$ across transform states); the jointly fitted
coefficients are collinearity-unstable, so interpretation rests on the
marginal analyses. In-family --- fitted and evaluated within the same
state pool --- the curve reaches $R^2 = 0.98$ over $32$ states
(Fig.~\ref{fig:ivf}), real and twin corpora landing on the same line --- the
statistics, not the corpus's identity, carry the prediction. The routing
term is causal, not merely fitted: the targeted probe of
Sect.~\ref{sec:learning} (iv) moves the routing margin alone and IVF
efficiency follows monotonically at every budget. The form's
deliverables are therefore the lever and the ordering --- which
statistic moves IVF, and which transforms help --- plus the sufficiency
evidence that licenses simulation; absolute recall levels for unseen
encoders are the twin's job (Sects.~\ref{sec:ladder} and \ref{sec:twin25k}).

\paragraph{Scope of the mechanism.} The derivation is about the
nearest-centroid routing decision, not about IVF's implementation, so its
qualitative content extends to any index whose candidate generation routes
through that decision: PLAID's token-level candidate generation (query
tokens routed to their nearest token-cluster centroids) and ScaNN's
partitioning stage face the same gap-survival problem, and the deployed
IVF-PQ composition is exactly this section's routing analysis times
Sect.~\ref{sec:pq}'s in-cell quantization analysis. This is stated as a
mechanism-scoped expectation, untested here: the sign and the lever --- the
routing margin, computed at the level the routing operates on (tokens for
PLAID) --- are expected to transfer; the fitted constants are not, and
PLAID's union of per-token routing decisions softens single-decision
failures, so levels and even orderings require measurement. The
measurement suite already records the token-level cluster panel, so the
expectation is directly testable with this paper's instruments.

\subsection{HNSW: score margins over cluster structure (derived by twin simulation)}\label{sec:hnsw}

A closed-form level predictor is a
\textbf{validated negative}: absolute-level forecasts
failed on the majority of held-out pairs in two successive evaluation
rounds (Sects.~\ref{sec:round1}--\ref{sec:round2}), which is
evidence that no formula over the measured statistics predicts HNSW recall levels ---
a result evaluated exactly as every positive claim was, and one the
theory literature makes expected: graph-search
guarantees exist only in restricted regimes (Sect.~\ref{sec:related} ---
low dimension, bounded intrinsic dimension), and worst-case constructions
for HNSW are provably bad. An independent attempt within this project ---
richer fitted geometry panels for IVF and HNSW, frozen before measurement
at one million documents --- also missed its accuracy target (IVF MAE
$0.176$ against $0.08$), corroborating that summary-statistic formulas do
not carry navigation levels; simulation does (Sect.~\ref{sec:twin25k}).
What replaces the formula is a two-factor account of the governing
statistics --- measured, in Sect.~\ref{sec:labels}'s sense --- plus a twin
simulator whose level predictions are derived in that same sense:
quantitative, under the listed cluster-mixture assumptions, run on the
twin.

\begin{figure}[t]
\includegraphics[width=\textwidth]{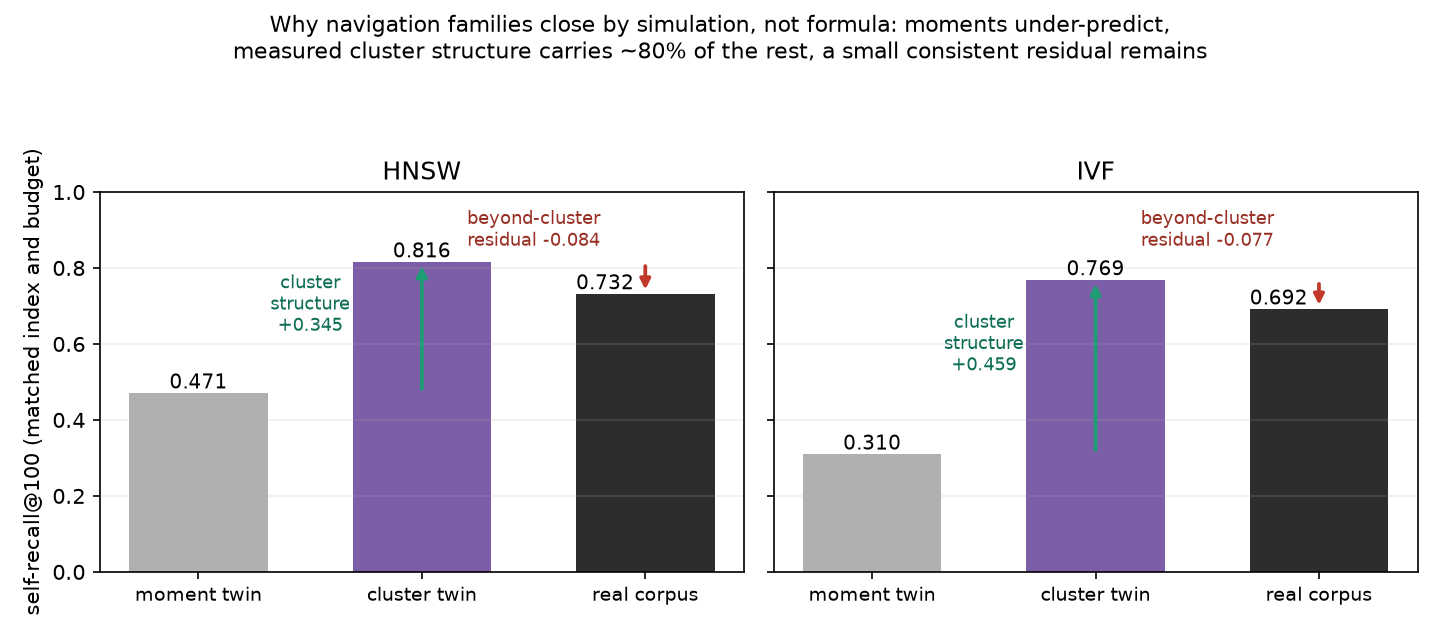}
\caption{Why the navigation families close by simulation, not formula.
Recall of the same index at the same budget on three versions of the same
corpus (BGE-large / MS MARCO): a moment-matched Gaussian twin, a
cluster-matched twin, and the real corpus. Moments under-predict both
families; measured cluster structure carries roughly $80\%$ of the
remainder; the small signed residual --- real corpora are slightly
\emph{harder} than their cluster-ideal --- is consistent across both
families and is the open quantity named in Sect.~\ref{sec:limitations}.}
\label{fig:navclosure}
\end{figure}

\emph{Across corpora and encoders}, graph-search efficiency tracks cluster
organization, while the intuitive alternatives fail measurably: hubness is
inconsistent (whitening lowers hub statistics yet lowers recall; the
moment twin has both the highest hubness and the lowest recall) and
moment-derived dimension statistics fail outright --- the moment twin shares
the real corpus's spectrum exactly (flatness $0.197$ both) yet loses
$0.26$ recall (Fig.~\ref{fig:navclosure}). \emph{Within a corpus}, encoder
ordering rides on score margins: the round-2 statistics order encoders at
$\rho = 0.706$ and $0.757$ on two independent held-out corpora
(Sect.~\ref{sec:round2}), and the Sect.~\ref{sec:learning} (iv)
dissociation makes the assignment causal --- sharpening margins lifts HNSW
($0.922 \to 0.994$ at efSearch $16$) while a $14\times$ routing-margin
increase leaves it flat. Anisotropy itself produces only a small
non-monotone dip ($0.02$--$0.04$ at intermediate concentration, at every
budget) --- second-order, and recoverable with compute.

The twin hierarchy turns this account into prediction
(Fig.~\ref{fig:navclosure}): moments alone under-predict HNSW recall by
$0.26$; adding measured cluster structure (the tolerance-checked cluster
twin) recovers $\approx 80\%$ of the gap; the signed remainder
($\approx -0.08$, consistent across both navigation families) is the named
beyond-cluster residual. Used as a zero-parameter simulator, the cluster
twin predicts absolute HNSW recall at MAE $0.039$ across $17$
encoders (Sect.~\ref{sec:twin25k}) --- a derived prediction in
Sect.~\ref{sec:labels}'s run-on-a-twin sense; the family closes by
simulation, with the closed-form failure as the contrast that makes the
closure meaningful.

\section{The Trade-off Calculus}\label{sec:calculus}

The pricing question is answered by composition. A corrective transform
enters the calculus twice: it spends fidelity --- predicted by the gap
model of Sect.~\ref{sec:fidelity} --- and it moves the statistics
Sect.~\ref{sec:families}'s map reads, which predicts the efficiency it
buys (closed forms for the fixed grids; twin simulation for navigation).
Sect.~\ref{sec:axes}'s bound, $r \ge p + s - 1$ at matched depth,
converts the two predictions into an end-to-end verdict before any index
is built. Fig.~\ref{fig:tradeoff}'s measured plane validates the
verdicts out of sample, and Sect.~\ref{sec:regimes} compresses them into
one operating rule.

\subsection{One pricing rule, at both extremes of the tax}\label{sec:regimes}

\begin{figure}[t]
\includegraphics[width=\textwidth]{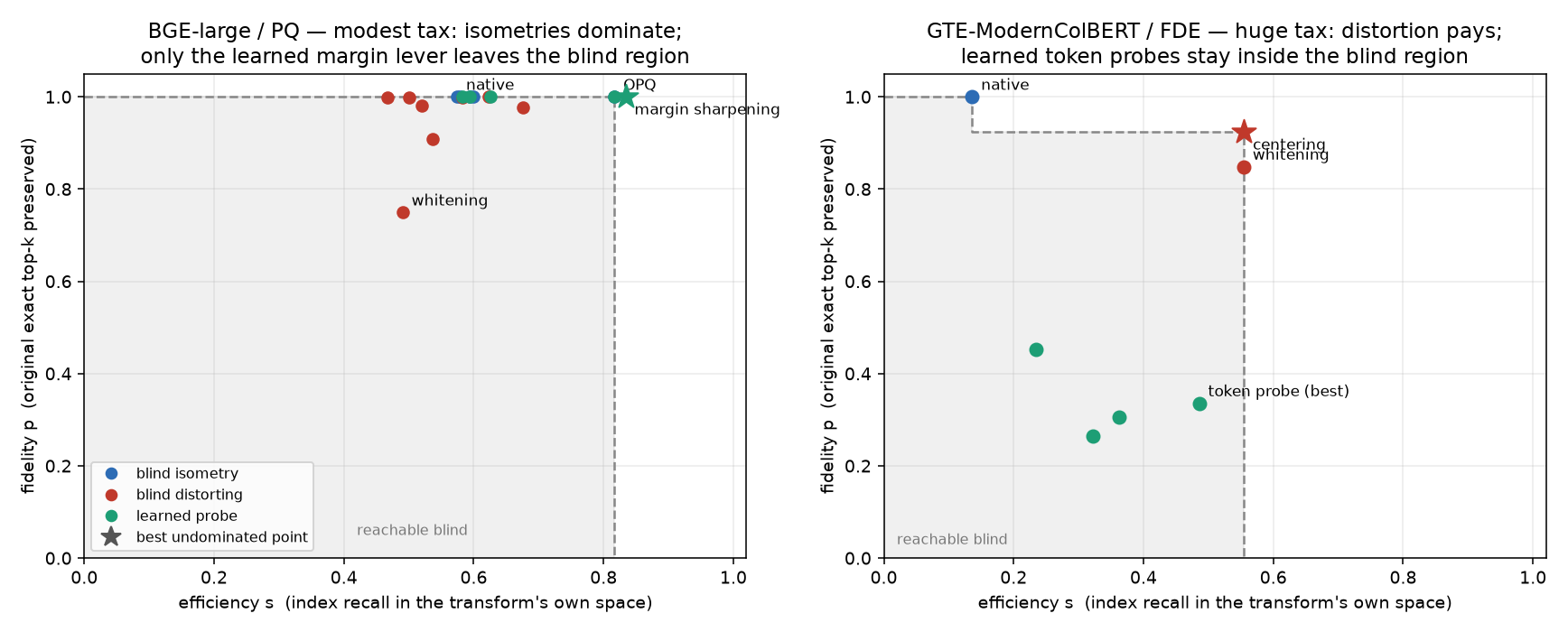}
\caption{The trade-off plane on real data. Every transform (blind and
learned) as a point in (efficiency, fidelity) for the dense/PQ pipeline
(left) and the late-interaction/FDE pipeline (right); isometries sit on the
$p=1$ line by construction.}
\label{fig:tradeoff}
\end{figure}

Served recall is lost through two channels: directly, the efficiency
the index loses to the corpus's geometry --- the tax --- and
indirectly, the fidelity a post-hoc correction spends while buying that
efficiency back. Optimizing end-to-end recall means striking the
balance between the two, and the balance is predictable --- down to the
cases where the right correction is none at all. The two measured
pipelines sit at opposite ends of the tax range;
Fig.~\ref{fig:tradeoff} plots every corrective transform on both as a
point in (efficiency, fidelity), and Fig.~\ref{fig:teaser}, left,
previews its two headline moves. On the dense corpus, where PQ's tax is
modest, every whitening strength loses to doing nothing ($0.534$ at the
sweep's best vs $0.575$ raw) while the zero-fidelity-cost OPQ rotation
wins outright ($0.817$): isometries dominate whenever they suffice. On
the late-interaction corpus, where FDE's tax is enormous, distortion is
worth its price: centering pays $7.6\%$ fidelity for a $3.4\times$
efficiency gain. One rule covers both: compare the efficiency gained
against the fidelity spent at the operating depth; rotations spend
nothing.

Family-specific verdicts, measured across all $22$ held-out
encoder--dataset pairs of the first evaluation round (Sect.~\ref{sec:round1}):
whitening damages graph search universally (all $22$ pairs,
median $-0.21$); centering is mild (helped $8$ of $22$, hurt none);
rotations are \emph{exactly} neutral for graphs and for IVF, as invariance
requires. For
late-interaction FDE, the centering decision itself is encoder-dependent
and forecastable (Sect.~\ref{sec:fderules}): centering \emph{hurts} low-MCS models (ColBERTv2:
$-0.12$ to $-0.16$) and rescues high-MCS ones ($+0.19$ to $+0.37$).

\subsection{Fidelity is predictable (derived under assumptions; promoted from measured on the held-out calibration of Sects.~\ref{sec:round1}--\ref{sec:round2})}\label{sec:fidelity}

A transform evicts a true neighbor only by moving score \emph{gaps}: with
per-triple gap perturbations measured on (query, top-$k$ item, boundary
item) samples, survival $\approx \Phi((\mathrm{margin} +
\mathrm{shift})/\mathrm{sd})$. Near neighbors perturb coherently --- the
naive independent-perturbation model over-predicts fidelity loss $3$--$4
\times$; the gap model calibrates at Spearman $1.00$, MAE $0.034$ in-family
and, on the $34$ fresh dense pairs of the second evaluation round
(Sect.~\ref{sec:protocol}), \textbf{MAE $0.008$} for centering (whitening: rank
$0.946$ with a documented optimistic bias at extreme distortion). The
mechanism it exposes: distorting transforms \emph{raise} the average gap while
adding gap noise; fidelity is lost to the noise, not the shift.

\section{Can the Correction Be Learned? Probes and the Margin Lever}
\label{sec:learning}

The learning question asked whether the correction can be learned with the
model rather than fitted to the corpus. Adapters on a frozen encoder join the transform pool as probes, under three
controls: a rank-only variant (no geometry objective --- the null
instrument), rotation-constrained variants (fidelity cost provably zero),
and a supervised variant kept separate (its gains ride on labels). Two
further probes each target a single statistic of the Sect.~\ref{sec:families} map: a
\emph{navigability} probe that directly optimizes the cluster routing margin
(shrink each document's distance to its own base-space centroid relative to
the nearest other, with rank and alignment terms holding the original
neighborhoods in place), and a \emph{token-level} probe for FDE that removes the
token cone under an alignment constraint. The probes train on frozen
encoders over the corpora of Sect.~\ref{sec:setup} --- (i)--(ii) on
BGE-large (in-sample $50$k MS MARCO shard), (iii)--(iv) on
bge-base-en-v1.5 (the qrel-bearing held-out $25$k shard), and the token
probe (v) on GTE-ModernColBERT ($25$k late-interaction shard) --- at the
training configurations of Appendix~\ref{app:details}. Throughout, a
probe's exact-neighborhood preservation is its fidelity $p$
(Sect.~\ref{sec:axes}), judged against a working preservation target of
$p \ge 0.95$. Five results.

\paragraph{(i) Fixed-grid geometry targets belong post-hoc:} even
dose-corrected --- matched in training strength --- gradient-trained
rotations reach MF $4.18$ vs the specialized OPQ solver's
$1.92$, with efficiency following the Sect.~\ref{sec:pq} curve --- the probe's failure
itself validates the map --- and the correction transfers across disjoint
shards (stale OPQ within $0.004$ of refit), so per-corpus refitting buys
nothing here.

\paragraph{(ii) The unique learned lever is margin sharpening:} the
rank-only adapter, trained label-free on exact neighbors, nearly
triples score margins and lifts efficiency for every family (PQ $+0.15$, IVF $+0.14$,
HNSW $+0.10$--$0.16$) at base fidelity $1.0$ --- no blind transform can mint
margins: rotations preserve every inner product and hence every margin,
and distorting transforms change margins only by changing the top-$k$
itself. Architecturally the probe is nothing but a trainable final encoder
layer, so this is an existence proof with training-time reach: two
encoders can agree on essentially every exact neighbor yet differ
$\sim\!3\times$ in margins --- and therefore materially in every index
family's recall at budget. A model's margins are a property encoder
training can target, not a datum to accept (Sect.~\ref{sec:discussion}).

\paragraph{(iii) It is not free in the task frame:} on the qrel-bearing
corpus (300 judged queries, 315 relevance pairs --- the paper's sparsest,
noisiest measurement), the cost is visible at exact scoring, before any
index exists: the probe preserves $0.9997$ of the encoder's top-$10$
within its $N{=}100$ operating candidate set and leaves exact qrel
recall@100 unchanged ($0.993$ in both spaces), but drops exact qrel
recall@10 from $0.992$ to $0.930$ --- the relevant documents survive among the exact candidates
but are reordered out of the top ranks, a relevance drift the
label-free frame cannot register (Sect.~\ref{sec:groundtruth}). At served budgets the same
drift surfaces as 3--5 points of task recall at the PQ and IVF operating
points
($0.935 \to 0.902$; $0.933 \to 0.883$, roughly 2--3 standard errors),
while on HNSW --- where the efficiency gain is largest --- task recall at
budget does not fall ($0.963 \to 0.980$ at efSearch 16, within
noise): the trade's sign depends on how much efficiency the lever buys
against the semantic shift it pays. What is general is that the trade
exists and must be audited --- the two-frame discipline of Sect.~\ref{sec:groundtruth} is what
catches it.

\begin{figure}[t]
\includegraphics[width=\textwidth]{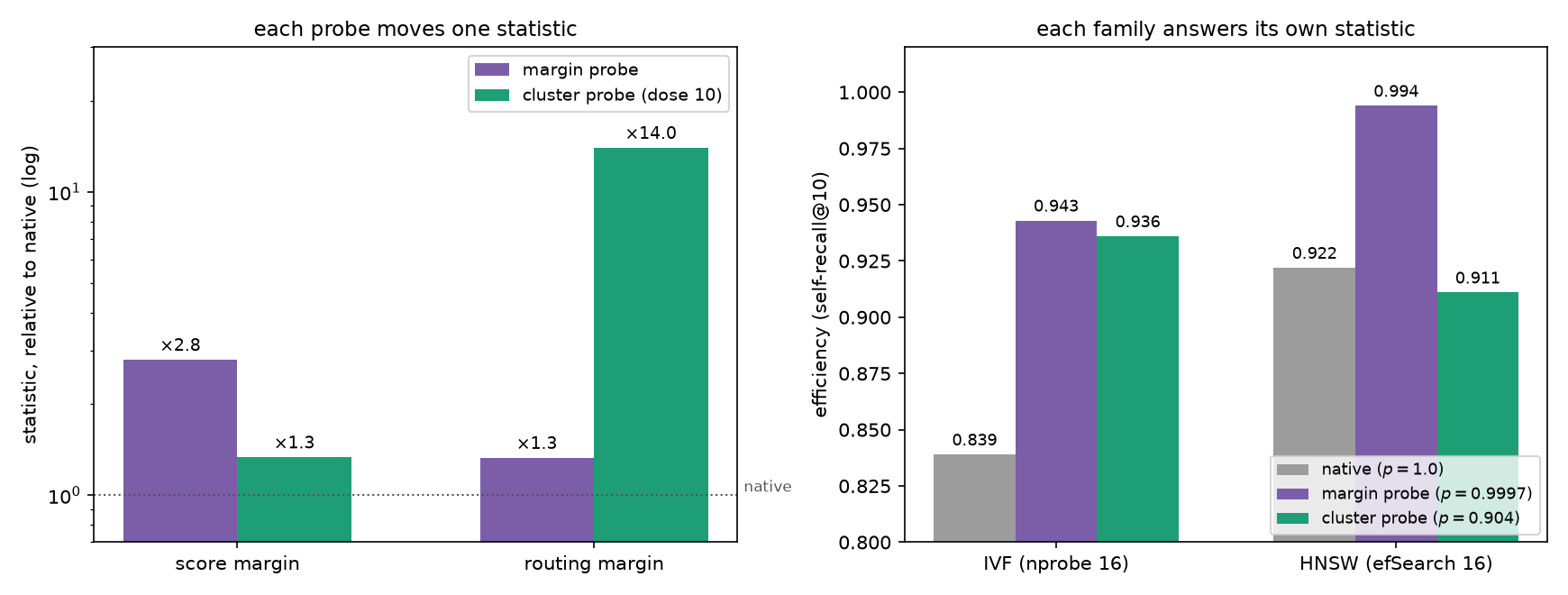}
\caption{The two-probe dissociation of the navigation families.
\emph{Left:} each probe moves essentially one statistic --- the
margin-sharpening probe of Sect.~\ref{sec:learning} (ii) sharpens score
margins ($\times 2.8$) while barely touching the routing
margin; the navigability probe of Sect.~\ref{sec:learning} (iv) moves
the routing margin ($\times 14$) while
barely touching score margins. \emph{Right:} IVF responds to both probes (its
sufficiency curve carries both statistics); HNSW responds only to the
margin-sharpening probe --- cluster separation above native levels leaves it flat.
Legend gives each state's exact-neighborhood preservation: the
margin-sharpening probe's gains come at $p \approx 1$, the navigability
probe's at $p = 0.904$.}
\label{fig:dissociation}
\end{figure}

\paragraph{(iv) The navigation frontier is real but not cheap --- and it dissociates
the two navigation families.} On the same qrel-bearing corpus as (iii),
the navigability probe moves the routing margin massively and
dose-responsively ($0.061 \to 0.853$ at the lower dose, $\to 1.692$ at the
higher) --- movement no blind transform can produce, since rotations leave
the routing margin invariant and the distorting transforms lower it --- but
fidelity breaks first: preservation falls to $0.904$ and $0.621$, below
the $0.95$ working preservation target, and the task price reaches $13$ points, far above
the margin lever's $3$--$5$. The map-consistency half of the test split into
a discovery. IVF efficiency follows the induced routing margin
monotonically at every probe budget ($0.656 \to 0.745 \to 0.799$ at nprobe
$4$, with the same ordering at $8$ and $16$); HNSW does not move ($0.922
\to 0.911 \to 0.908$ at efSearch $16$) --- yet the margin-sharpening probe of
(ii), which nearly triples score margins while moving the routing margin only
$+34\%$, lifts HNSW to $0.994$ at that same budget. Two probes
(Fig.~\ref{fig:dissociation}), each moving one statistic: each navigation family responds causally to exactly the
statistic its Sect.~\ref{sec:families} predictor reads --- a confirmation of the per-family
assignment that correlation alone could not give. This does
not contradict the twin decomposition of Sect.~\ref{sec:twin25k}, where \emph{removing} the
corpus's cluster structure costs HNSW heavily: the probe moves separation
\emph{above} native levels, a region where HNSW's graph search is already
saturated on this corpus, while IVF's partition routing keeps gaining.

\paragraph{(v) Post-hoc wins for FDE too --- tested, not assumed.} FDE is the one
family whose post-hoc correction measurably costs fidelity (centering,
$p = 0.924$), so a learned alternative had its best case here. The
token-level probe removes the cone (token MCS $0.910 \to 0.004$) but loses
to blind centering on \emph{both} axes: fidelity $p\,0.27$--$0.45$ against
$0.924$, candidate recall $0.10$--$0.17$ against $0.468$: a clear failure. The mechanism is instructive: centering is a uniform shift, so
relative token geometry survives it; with a cone this tight (MCS $0.91$)
every token must move far during any cone removal, so a per-token map's
alignment term carries no ranking signal, and the learned map scrambles
exactly what the uniform shift preserves. Within the probe's dose sweep the
map's direction still holds --- recall rises as MCS falls ($0.110 \to 0.116
\to 0.168$ as MCS drops $0.196 \to 0.018 \to 0.004$).

\section{The Forecast Protocol and Results}\label{sec:evaluation}

\subsection{Evaluation protocol}\label{sec:protocol}

All formulas and models were developed on the in-sample corpus; every
result in this section is evaluated out-of-sample, on encoder--dataset
pairs that played no role in development. Two evaluation rounds were
run: 22 pairs across 11 encoders, then --- after the round-1 revisions ---
34 dense and 6 late-interaction pairs on fresh encoders and datasets;
the twin-simulation tests and the two targeted-probe tests of
Sect.~\ref{sec:learning}~(iv)--(v) follow the same discipline. The FDE rules were additionally
evaluated prospectively on a corpus selected only after every rule,
threshold, and seed had been fixed (Sect.~\ref{sec:fderules}). Read against the
three questions: Sects.~\ref{sec:round1}--\ref{sec:round2} test the prediction question on the fixed-grid
families and record the navigation families' closed-form failures; Sect.~\ref{sec:fderules}
tests the FDE rank rules; Sect.~\ref{sec:twin25k} closes the navigation families by twin
simulation and ablates which inputs carry the prediction --- corpus
statistics suffice, and the query distribution is the residual axis;
Sect.~\ref{sec:milliondoc} takes the twin to one million documents on an unseen corpus. One
global check spans every round: across every family and state measured
in this paper ($116$ HNSW rows included), the exchange bound of
Sect.~\ref{sec:axes} is violated zero times.

\subsection{Fixed-grid forecasts transfer; navigation formulas fail}\label{sec:round1}

\begin{figure}[t]
\includegraphics[width=\textwidth]{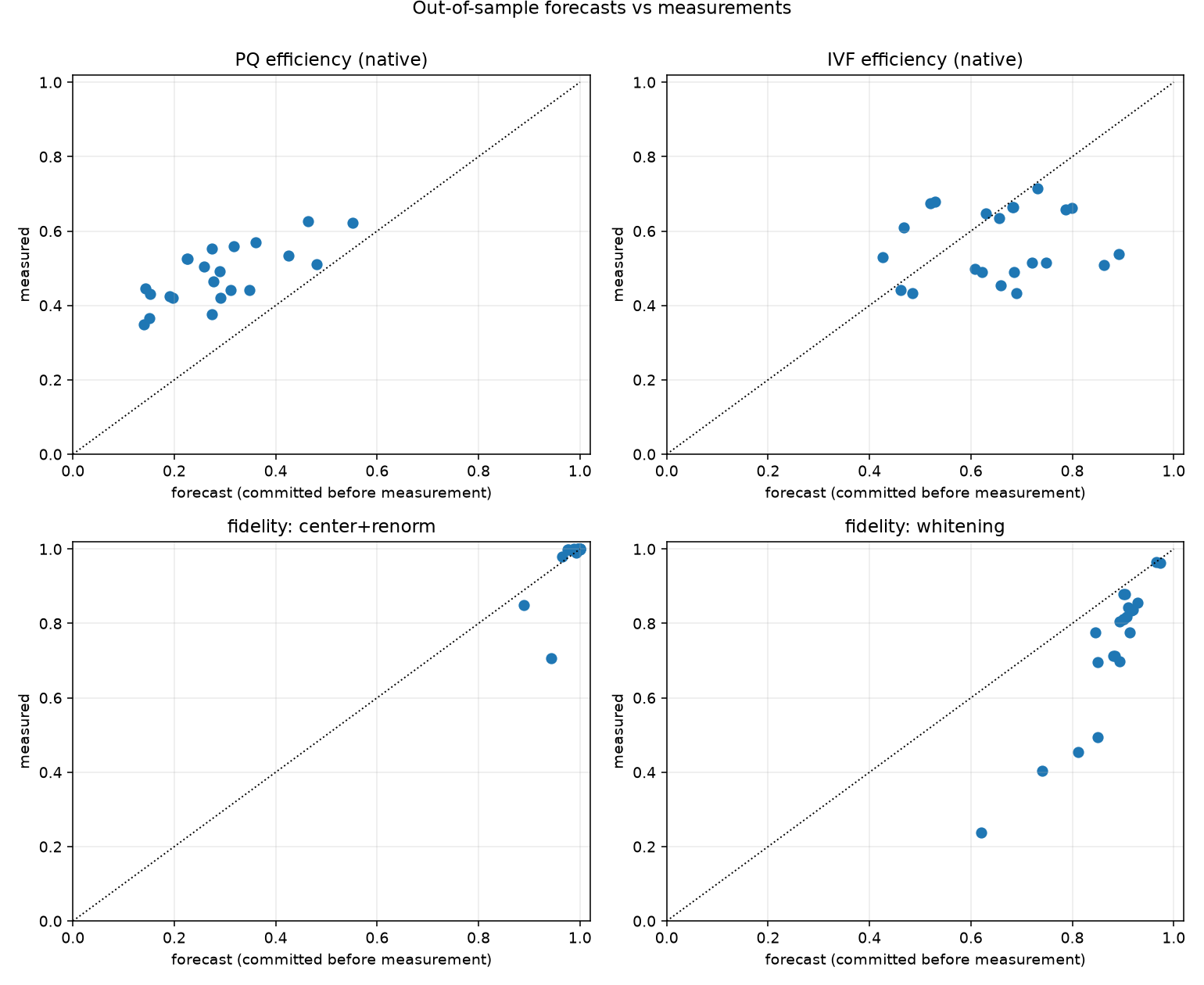}
\caption{Forecast versus measured recall on held-out pairs (round 1
shown; round-2 scorecards in the artifact). Each point is one
held-out encoder--dataset pair.}
\label{fig:scorecard}
\end{figure}

In the first round (22 held-out pairs; Fig.~\ref{fig:scorecard}), fidelity forecasts calibrated
immediately (centering MAE $0.019$). The
misalignment factor forecast the \emph{decision} (does OPQ help?) on $20$ of
$22$ pairs and
within-corpus ordering at $\rho = 0.86$; absolute PQ levels missed (MAE
$0.19$) --- diagnosed not as missing physics but as fit-population mismatch,
and the refit on encoder-native states transfers at leave-one-encoder-out
MAE $0.050$. IVF's curve missed ordering; the recorded-but-unused cluster
excess restores it ($\rho\, 0.42 \to 0.76$). HNSW's absolute-level
intervals held on only $11$ of the $22$ pairs. \textbf{Takeaway: closed forms
carry the fixed-grid families; navigation levels resist them.}

\subsection{The revised models on fresh encoders and datasets}\label{sec:round2}

In the second round (34 fresh dense pairs), with the revised models
held fixed: PQ MAE $0.089$, rank $0.900$, \textbf{the
OPQ-helps sign correct on all $34$ pairs} (with the post-OPQ misalignment now predicted, not assumed);
fidelity-center MAE $0.008$; IVF cross-corpus levels MAE $0.105$, rank
$0.779$ --- while within-corpus encoder ordering stayed weak ($\rho\,
0.07$--$0.21$), a documented negative. HNSW absolute levels failed again ($14$ of $34$ within interval) ---
while the same numbers order encoders within each corpus at
$\rho = 0.706$ and $0.757$ (a post-hoc rank analysis, replicated
independently on both corpora). The
strictest validity cut --- the six encoders appearing in no fit anywhere ---
\emph{strengthens} the results (PQ rank $0.958$, sign correct in all $12$,
fidelity MAE $0.008$). \textbf{Takeaway: the fixed-grid forecasts transfer to
fresh encoders and datasets; unanchored HNSW levels do not --- the
negative that motivates the twin (Sect.~\ref{sec:twin25k}).}

\subsection{The FDE rules, across encoders and prospectively}\label{sec:fderules}

Both rank rules held on all $12$ ordered pairs: raw FDE recall
descends in token MCS, and centering recovery ascends in it --- including
the sign result that centering harms the low-MCS encoders. Collision rates
ordered correctly ($\rho = 0.886$); their absolute levels over-predict
(synthetic calibration population), stated as ordering-grade.

\paragraph{A prospective confirmation on an unseen corpus.} A third evaluation
tested the FDE rules prospectively --- every rule, threshold, and seed
fixed before the evaluation corpus was resolved --- on a corpus no
earlier round touched: BEIR HotpotQA, four late-interaction encoders (a
corpus-transfer test; the encoders had appeared in development). The
MCS direction ordered native FDE self-recall (the efficiency $s$ of
Sect.~\ref{sec:axes}) on five of six pairs ($\rho = -0.800$), and an MCS threshold of
$0.26$, fixed during development, predicted all four centering signs
(macro mean absolute effect $0.073$) --- including the harm side,
ColBERTv2 at $-0.023$. The measured exchange replicates Sect.~\ref{sec:fde}'s account: centering
preserved $0.598$--$0.842$ of the native top-$100$, and whitening spent
more fidelity in every case --- intensifying the gain for
the high-MCS encoders, dominated for the low-MCS ones. With four encoders
the ordering evidence is directional (a uniform rank null gives
probability $1/6$). \textbf{Takeaway: one statistic read from raw token
embeddings decides FDE behavior --- direction, magnitude, and the
centering decision --- including prospectively on a corpus never used in
development.}

\subsection{Twin simulation closes the navigation families}\label{sec:twin25k}

The cluster-matched twin meets its matching tolerance (cluster panel
within $2\%$) and
decomposes its reference corpus (BGE-large / MS MARCO, Fig.~\ref{fig:navclosure})
exactly: HNSW $0.732 = 0.471$ (moments) $+
0.345$ (cluster structure) $- 0.084$ (beyond-cluster residual); IVF $0.692
= 0.310 + 0.459 - 0.077$. Cluster structure carries $\approx 80\%$ of the
beyond-moment behavior of both navigation families, and the residual ---
real corpora are slightly \emph{harder} than their cluster-ideal --- is
consistent across them.

Used as simulators (zero fitted parameters; every level claim is held
to a working accuracy target of MAE $\le 0.05$), the twins close what
closed-form statistics could not: \textbf{HNSW absolute recall, MAE $0.039$
(bias $+0.035$); IVF within-corpus levels, MAE $0.034$ (bias
$+0.033$)} across $17$ held-out encoders at the $25$--$50$k scales
studied --- build the twin, not the index. The known bias equals the beyond-cluster
residual (over-prediction, largest on the two Arctic encoders; roster
in Appendix~\ref{app:details}). A cheaper
practical alternative for HNSW --- measure one encoder, predict the rest
by applying its offset --- achieves median MAE $0.032$--$0.036$ robustly over
every anchor choice.

\paragraph{What the statistics carry: an ablation of the prediction question's
claim.} The prediction question asked whether recall can be forecast from label-free
statistics alone, and for the navigation families the affirmative answer
runs through the twin --- which is nothing but the cluster panel made
generative: the structure that inter/intra-cluster distances and the
routing margin summarize is fitted (cluster weights, centroids,
within-cluster covariances) and sampled, and the matching tolerance is
stated in the panel's own metrics. The analyses below ablate that claim
against the strongest empirical alternative and along each input axis;
rows marked post-hoc in Table~\ref{tab:fullcorpus} characterize the instrument and change
no headline result.

\paragraph{The corpus side saturates.} If the corpus is in hand there is a simpler
predictor than any statistic: build the target index on small uniform
samples and extrapolate recall against log effective fraction. Run on the
same 17 encoders (Table~\ref{tab:fullcorpus}), this
baseline is the better HNSW level predictor, while its IVF levels fail ---
the target's $1024$ cells cannot exist on a few-hundred-document sample ---
and the twin, simulating at the target's own settings, keeps IVF MAE
$0.033$ (a three-seed re-mint of the default configuration; the
single-draw instrument of Table~\ref{tab:fullcorpus} reads $0.034$, within its stated draw
agreement). The decisive comparison is at matched inputs: twins minted at
the subsample's own sizes and queries predict indistinguishably from real
subsamples at every scale (HNSW MAE within $0.003$ at all four sizes, in
successes and failures alike; Appendix~\ref{app:details}). Replacing the real corpus by a
draw from its cluster statistics costs nothing --- the statistics named by
the prediction question contain everything these indexes consume from the
corpus side.

\paragraph{The query side is a distribution-matching problem under a scale
asymmetry.} The corpus funds its statistics --- $25{,}657$ samples fit $64$
clusters with covariances --- while a $1{,}000$-query sample funds neither
its own moments nor, it turns out, useful traffic weights. Holding the
twin corpora fixed and varying only what the simulation knows about the
query distribution gives a ladder that is monotone in distribution match
(Appendix~\ref{app:details}): a ridged Gaussian fitted on the real queries fails outright
(HNSW MAE $0.207$ --- two moments under-describe the query distribution
exactly as they under-describe corpora); reweighting the corpus mixture
by query traffic over the $64$ clusters regresses below the default
document-proxy ($0.060$ vs $0.040$) --- queries already touch $63$ of $64$
clusters, so the traffic marginal was never the mismatch, and fixing one
marginal while keeping the wrong within-cluster conditional leaves the
actual mismatch in place; the real queries
close the gap ($0.012$). The residual mismatch is itself measurable in
advance: queries sit at cluster fringes (mean cosine to own centroid
$0.659$ against the documents' $0.708$), a scalar a small query sample
funds --- the query side's analogue of the corpus tolerance.

\begin{table}[t]
\caption{Predicting full-corpus recall ($25{,}657$ documents, $1{,}000$
queries, $17$ encoders) without building the full index. MAE = mean
$|$predicted $-$ measured$|$ self-recall@100 over the $17$ encoders;
bias = the mean signed error, predicted minus measured (positive =
over-prediction); bold marks the best MAE per column. Rows marked
$\dagger$ are post-hoc instrument analysis; the twin row is the default
configuration, single draw (three-draw averages agree within $0.007$).
Per-fraction subsample readouts and the query-source ladder are in
Appendix~\ref{app:details}.}
\label{tab:fullcorpus}
\centering
\footnotesize
\setlength{\tabcolsep}{3pt}
\begin{tabular}{@{}p{3.3cm}p{2.9cm}p{1.6cm}rr@{}}
\toprule
predictor & documents & queries & HNSW MAE (bias) & IVF MAE (bias) \\
\midrule
cluster twin --- statistics only, the prediction-question instrument & 25,657 synthetic (from statistics of all 25,657 real) & 1,000 document-proxy & 0.039 (+0.035) & 0.034 (+0.033) \\
cluster twin + real query log $\dagger$ & 25,657 synthetic & 1,000 real & \textbf{0.011} ($-$0.011) & \textbf{0.027} ($-$0.023) \\
twin, subsampled + extrapolated (control) $\dagger$ & 513--5,131 synthetic $\times$ 3 draws & real & 0.013 ($-$0.003) & 0.233 ($-$0.233) \\
twin, subsampled + extrapolated (control) $\dagger$ & 513--5,131 synthetic $\times$ 3 draws & document-proxy & 0.030 (+0.016) & 0.196 ($-$0.196) \\
subsample of the real corpus, extrapolated --- empirical control, no statistics & 513--5,131 real $\times$ 3 seeds & real & 0.012 (+0.002) & 0.244 ($-$0.244) \\
\bottomrule
\end{tabular}
\end{table}

\textbf{Takeaway: at corpus scale, the cluster statistics contain everything
the navigation indexes consume from the corpus side; the residual error
is query-distribution mismatch, measurable in advance.}

\subsection{The million-document scale test}\label{sec:milliondoc}

Both query configurations of
the full-size twin were then run at $1$M on an unseen corpus (HotpotQA; four dense encoders), against
target measurements produced independently and read only after all
simulation outputs were written. The verdict splits. \textbf{IVF absolute
levels transfer: the real-query twin reaches MAE $0.027$ at $1$M (bias
$+0.022$)} ---
and outperforms every alternative tested on the same targets, including
a subsampled-real-index baseline ($0.046$) and fitted geometry panels
($\ge 0.059$). \textbf{HNSW absolute levels do not transfer for the full-size construction} (statistics-only twin MAE $0.124$, real-query twin $0.204$; both fail): a validated
negative with a mechanism --- the signed bias \emph{flips} from $+0.03$
over-prediction at $25$--$50$k to $0.12$--$0.20$ \emph{under}-prediction at
$1$M, so the beyond-cluster residual is scale-dependent: at $1$M the
real corpus carries fine-grained local structure that aids graph
navigation, and the twin's within-cluster Gaussians erase it.

After removing each
configuration's constant offset, per-case dispersion is $0.017$--$0.022$
for every configuration and family: the $1$M prediction error is almost
entirely a single reproducible offset, not noise. Three of four encoders also exceed the $2\%$ matching
tolerance at $1$M --- direct evidence, independent of any prediction,
that the fixed $k{=}64$ recipe under-resolves million-document cluster
structure --- though meeting the tolerance alone is not sufficient for
level accuracy. The twin's absolute-level
claims are therefore scoped: both families at the corpus scales
validated above; at $1$M, IVF levels (real-query twin), with every other
configuration's error a reproducible offset ($-0.06$ to $-0.20$) over
$\approx 0.02$ dispersion.

\paragraph{The size ladder (Sect.~\ref{sec:twin}).} Fitting
the statistics on only
$25{,}600$ sampled documents ($2.6\%$ of the corpus), minting twins at
$3{,}200$--$25{,}600$ documents, and extrapolating recall against log
corpus size to $1{,}001{,}595$ recovers most of what the full-size twin
loses on HNSW: MAE $0.057$ with real queries (statistics-only queries:
$0.094$). The extrapolation rides the
size trend measured at scales where the twin is known to track reality,
instead of building the under-resolved million-document twin. The ladder was run
as a follow-up to the full-size failure and confirmed by a replication
with fresh randomness (MAE $0.064$; the two runs agree within $0.01$,
and neither meets the $0.05$ working target unanchored). The ladder's error is a
reproducible constant offset ($-0.057$ and $-0.064$ across the two
independently seeded runs) over per-case dispersion $\approx 0.02$ ---
which makes it calibratable, and anchoring confirms it at $1$M: taking
each encoder in turn as the one measured anchor (its mean prediction
error over three budgets, subtracted from the other three encoders'
predictions), the median MAE across the four anchor choices is $0.027$
(per-anchor $0.022$--$0.034$) in run 1 and $0.026$ ($0.022$--$0.031$) in
the replication --- inside the $0.05$ working target the unanchored
ladder misses, and matching the $25$--$50$k precedent. The residual
above pure dispersion reflects offset variation across anchors
($-0.033$ to $-0.087$), so the result is labeled what it is ---
statistics plus one anchored measurement, not statistics alone. Being the more
statistically disciplined construction (multi-draw, multi-size), the
ladder is the twin's recommended operating mode for level prediction at
scale --- anchored where the $0.05$ working target must be met --- with full-size
minting retained for attribution and for hosting the target's true cell
structure (which is what wins IVF). As with the closed-form failures
before it, the negative result localizes to the construction, not the
statistics: used within their resolution limit, the same cluster
statistics still carry the prediction. Its IVF
rows are the floored-cells control (cell counts cannot scale
below $16$ at these sizes; predictions clip). Table~\ref{tab:million} collects every
$1$M predictor with its document accounting; the two twin modes cover
each other's failure --- the full-size twin hosts the target's true cell
structure (best IVF predictor at $1$M), the ladder avoids the
under-resolved full-size build (HNSW from a $25{,}600$-document budget) ---
and with the corpus itself in hand, the subsample remains the best HNSW
level predictor.

\begin{table}[t]
\caption{Predicting recall at one million documents (unseen corpus,
$1{,}001{,}595$ total documents, $1{,}000$ queries, $4$ encoders;
targets measured independently, read only after simulation outputs were
written). Real documents consumed vs documents in the built index are
stated per predictor. Bold marks the best MAE per column among the
statistics-based predictors; the last row requires the corpus itself
and is the empirical control. $\dagger$: follow-up analyses to the
full-size result; the real-query ladder row is replicated with fresh
randomness. The anchored row applies the leave-one-encoder-out offset
correction, reported as the median across the four anchor choices.}
\label{tab:million}
\centering
\footnotesize
\setlength{\tabcolsep}{2pt}
\begin{tabular}{@{}p{2.4cm}p{2.0cm}p{2.0cm}p{1.1cm}p{2.4cm}p{1.6cm}@{}}
\toprule
predictor & real documents consumed & index built on & queries & HNSW MAE (bias) & IVF MAE (bias) \\
\midrule
full-size twin, real queries & 1,001,595 (statistics only) & 1,001,595 synthetic & 1,000 real & 0.204 ($-$0.204) & \textbf{0.027} (+0.022) \\
full-size twin, statistics-only queries & 1,001,595 (statistics only) & 1,001,595 synthetic & 1,000 proxy & 0.124 ($-$0.124) & 0.092 (+0.092) \\
ladder twin, real queries $\dagger$ & 25,600 (statistics only) & $\le$ 25,600 synthetic, 4 sizes $\times$ 3 draws & 1,000 real & \textbf{0.057} ($-$0.057); replication 0.064 ($-$0.064) & floored-cells control \\
ladder twin, statistics-only queries $\dagger$ & 25,600 (statistics only) & $\le$ 25,600 synthetic & 1,000 proxy & 0.094 (+0.094); replication 0.085 (+0.085) & floored-cells control \\
ladder twin, real queries + one anchored encoder & 25,600 (statistics) + one measured encoder (3 budgets) & $\le$ 25,600 synthetic & 1,000 real & 0.027 (+0.001); replication 0.026 ($-$0.002) & out of scope (full-size twin passes unanchored) \\
subsampled real corpus, extrapolated (same targets, run alongside the scale test) & 20,032--201,921 (built directly) & real subsets & 1,000 real & 0.007 (+0.003) & 0.046 (+0.046) \\
\bottomrule
\end{tabular}
\end{table}

\paragraph{Division of labor.} With only corpus statistics, the twin is
the instrument, and a strong default --- its document-proxy queries beat
both statistics-funded alternatives. With a query log as well --- the
common production case --- keep the twin corpus and use the real queries:
at the $25$k scale this gives subsample-grade HNSW accuracy without the
corpus and IVF levels nine-fold better than any subsample (Table~\ref{tab:fullcorpus}); at
$1$M it gives the best IVF levels of any predictor tested and --- with
one measured encoder as anchor --- HNSW within $0.03$ (Table~\ref{tab:million}).
With the corpus itself in hand, sampling it beats the statistics only on
HNSW levels --- a tie at $25$k, a clear win at $1$M. For attribution ---
decomposing behavior into layers of structure --- the twin is the only
instrument of the two. \textbf{Takeaway: at one million documents, statistics
alone carry partition indexes; statistics plus one calibration
measurement carry graphs.}

\subsection{Robustness}\label{sec:robustness}

Every reported family was re-run at $3$ index-training seeds: over $39$
reported quantities, median seed deviation $0.000$, $90$th percentile
$0.004$, maximum $0.012$ --- effects exceed seed noise by $5$--$100\times$.
Budget stability: geometry effects on HNSW preserve encoder ordering
across efSearch $16$--$128$ (rank agreement $0.82$--$0.96$), and the
synthetic non-monotone dip persists at every budget while shrinking with
it.

\section{Discussion}\label{sec:discussion}

\paragraph{A two-mechanism map.} Fixed-grid indexes are governed by formula-ready
statistics --- a mean-direction term for hashing, a covariance-block term for
quantization, margins for tolerance --- and close through closed-form
prediction. Navigation indexes (partitions and graphs) consume cluster
organization itself, which formulas summarize imperfectly but a
cluster-matched twin reproduces; they close through simulation. The
closed-form failures for navigation families are not embarrassments but
the contrast that makes the twin result meaningful.

\paragraph{The map covers the deployed landscape.} Production systems are
compositions of exactly the mechanisms analyzed: IVF-PQ is routing
survival (Sect.~\ref{sec:ivf}) times in-cell grid distortion (Sect.~\ref{sec:pq}); ScaNN composes the
same two stages; PLAID composes token-level routing with residual
compression; and a MUVERA pipeline composes sketch collision (Sect.~\ref{sec:fde}) with
whatever dense index searches the FDE vectors --- a second pass through the
map. Because each analysis attaches to the decision a stage makes, not to
the implementation that makes it, the per-stage audit reads onto these
systems directly, as mechanism-scoped expectations (Sect.~\ref{sec:ivf}) --- signs and
levers transfer, constants require fitting. And the twin needs no such
caveat: it is a corpus replica, not a per-index model, so \emph{any} index ---
composed, tuned, or not yet invented --- can be simulated on it before
being built. The practical rule has three
tiers (Sect.~\ref{sec:milliondoc}): with statistics alone, twin it; with a query log, keep
the twin corpus and the real queries; with the corpus in hand, sampling
adds only HNSW levels.
The methodology outlives the four families measured here.

\paragraph{Where corrections belong.} The demonstrated direction is
training time. A post-hoc correction buys its efficiency at two prices:
the fidelity a distorting transform spends
(Sect.~\ref{sec:calculus}), and an operational one --- a corpus-fitted
transform must be fitted, stored, and refit as the corpus changes. The
margin probe shows the decisive statistic can instead be shaped at the
source: a trainable final layer --- the cheapest possible form of
``more training'' --- produced an encoder with essentially identical
exact neighbors ($p \approx 1$), $2.8\times$ the margins, and
materially higher recall at every index budget, and it leaves no
corpus-fitted state in the serving path. For the navigation families
this extends the reachable frontier outright: no blind transform can
mint margins (Sect.~\ref{sec:learning} ii), and
Sect.~\ref{sec:regimes}'s verdicts leave those families no positive
post-hoc lever, so the learned members of Fig.~\ref{fig:tradeoff} sit
where no post-hoc point can. And because the target is a statistic
rather than an index, the move is not tied to any one structure:
Sect.~\ref{sec:families} establishes quantitatively that PQ, IVF, and
HNSW each read the margin --- it enters Sect.~\ref{sec:pq}'s statistic,
Sect.~\ref{sec:ivf}'s score-margin term, and Sect.~\ref{sec:hnsw}'s
assignment --- so a single trained encoder shifts the trade-off
frontier for every index family that depends on the property at once,
and the measured lifts land on all three simultaneously
(Sect.~\ref{sec:learning} ii). Index-friendliness is therefore a
trainable property of an embedding model, not a datum to accept from
the encoder. The scope is measured in both directions: for the
fixed-grid geometry targets the verdict reverses --- specialized
post-hoc solvers beat learned maps at reaching target geometry, and
their corrections transfer across shards of a distribution
(Sect.~\ref{sec:learning} i), so those corrections stay post-hoc ---
and the margin lever is label-free, so it can only sharpen the frozen
encoder's own opinion of relevance: at exact scoring it demotes
top-rank relevance (qrel recall@10 $0.992 \to 0.930$) --- a
3--5-point served-recall price at PQ and IVF budgets
(Sect.~\ref{sec:learning}, iii). Whether \emph{joint} training ---
a margin term alongside the task objective, where relevance and
separability are optimized together rather than sequentially --- escapes
that price is the natural next question. The machinery is established:
Poeem, JPQ, JTR, and EHI already co-train encoders with specific indexes
under supervision (Sect.~\ref{sec:related}); what those systems lack is a transferable,
index-agnostic target, which is precisely what the margin statistic
provides. Full encoder training is beyond this paper's scope, and the
hypothesis stands as the open question the existence proof motivates,
not as a result. Operationally, the stakes concentrate
in the navigation families: they typically serve corpora under continuous
change, and they are precisely the families for which post-hoc transforms
offer almost nothing to move (Sect.~\ref{sec:regimes}: rotations exactly neutral, whitening
harmful on all 22 pairs) --- a margin-trained encoder carries its
index-friendliness to every corpus it serves, with no per-corpus artifact
to fit, store, or refit.

\paragraph{Predict, train, deploy.} The economics of Sect.~\ref{sec:ladder}'s procedure are
asymmetric: the statistics pass costs seconds per encoder and twin
simulation minutes, while a single full $1$M HNSW build costs minutes
and the complete measured grid hours (Appendix~\ref{app:details}) --- every verdict is
available before anything is built. The margin result adds the second
leg: where the map says no post-hoc lever exists, the statistic the
serving index needs is trained into the encoder (Sect.~\ref{sec:learning}). What deploys
carries no per-corpus correction machinery --- nothing to fit, store, or
refit as the corpus changes.

\section{Limitations}\label{sec:limitations}

The cone model idealizes; Theorem~\ref{thm:degeneration}'s constants are loose. The full twin
decomposition is direct on one corpus (with indirect replication through
the forecast transfer of twin-derived statistics across $40$ pairs, and
the simulator validated on a second corpus). FDE breadth is three fresh
encoders --- ordering-grade claims only; an absolute FDE level model is open.
Scalar/binary quantization is not analyzed here; its designated
statistics (variance balance, kurtosis) are recorded but unused. HNSW ordering and the
anchored scheme are post-hoc analyses; unanchored
closed-form HNSW levels are a validated negative, not an open question.
The margin lever's task-frame price is measured on one corpus with
sparse labels (315 judged pairs), as are the navigability probe's. The
Sect.~\ref{sec:twin25k} instrument ablation (document parity, the query-source ladder) is
post-hoc analysis on one corpus family at $25$k scale. The $1$M test
promoted the real-query twin for IVF levels; HNSW absolute levels via
the full-size construction are scoped to the $25$--$50$k scales
validated, with the size ladder carrying $1$M HNSW at
$\approx 0.06$ unanchored (above the $0.05$ working target) and $0.026$--$0.027$
given one anchored measurement; the scale-dependence of the
beyond-cluster residual (its sign flips between $25$k and $1$M) remains
open, and the unmet matching tolerance at $1$M names the natural next
instrument --- a scale-aware twin whose cluster count grows with the
corpus, preserving the local structure a fixed-resolution recipe
smooths away --- which is beyond this paper's scope. Learned probes are bounded by the training budget used here,
and the FDE token probe's loss to centering is a verdict on
alignment-constrained cone removal --- a ranking-supervised token objective
(e.g., MaxSim-contrastive) is untested. The beyond-cluster residual
($\approx 0.08$) and closed-form derivations for the navigation families
remain open problems. The full measurement grid tops out at $50{,}000$
documents; the million-document evidence is one corpus and four dense
encoders (Sect.~\ref{sec:milliondoc}).

\section{Conclusion}\label{sec:conclusion}

Approximate retrieval behaves predictably --- but only if the right
statistics are measured and the right instrument is matched to each index
family. Formulas govern the fixed grids; cluster structure, measured by
inter/intra distances and reproduced by matched twins, governs navigation;
margins price every correction. The map was validated out-of-sample ---
every forecast tested on encoders and corpora withheld from
development --- and each family closed through the instrument it
demanded: where a closed form provably
fails, the failure itself points to the instrument that succeeds. The
result is an audit a
practitioner can run in one pass over raw embeddings, and a hierarchy of
synthetic twins that turns ``which structure matters'' from a debate into a
measurement. Because the analyses attach to mechanisms rather than
implementations, the audit extends stage-by-stage to the deployed
compositions of them --- IVF-PQ, ScaNN, PLAID --- and the twin is
index-agnostic: whatever index comes next can be simulated on a corpus's
statistical replica before anyone builds it. The practical summary is the
paper's workflow --- predict, train, deploy: forecast index
behavior and correction pricing from statistics and twin simulation,
train the margins the serving index needs into the encoder, and deploy
without per-corpus correction machinery.

\paragraph{Reproducibility.}

All numbers and figures in this paper regenerate from the
supplementary artifact, which contains the measurement and simulation
code, its configuration, and per-run manifests.
Embeddings rebuild from public models and datasets. A single Python
environment suffices; CPU covers everything except encoding.

\appendix
\section{Formal Statements and Proof Sketches}\label{app:proofs}

\paragraph{The $\rho$-cone model.} Fix a unit vector $u \in \mathbb{R}^d$ (the
cone axis) and $\rho \in [0,1)$. Every token embedding takes the form
$x = \rho u + \sqrt{1-\rho^2}\,z$, where the residual $z$ is a unit
vector orthogonal to $u$, drawn isotropically in $u$'s orthogonal
complement; all vectors share the same $u$ and $\rho$. Consequently
$\|x\| = 1$, and the measured mean cosine similarity has expectation
$\rho^2$ and concentrates on it (cross-residual terms vanish in
expectation), so we write $\mathrm{MCS} = \rho^2$ throughout.

\begin{lemma}[exact ranking is cone-invariant]\label{lem:invariance}
In the $\rho$-cone model, exact inner-product and MaxSim rankings are
independent of $\rho$.
\end{lemma}
\smallskip\noindent\emph{Sketch.} For any query token $x_q$ and document token $x_d$,
orthogonality of residuals to $u$ gives
$\langle x_q,x_d\rangle=\rho^2+(1-\rho^2)\langle z_q,z_d\rangle$. The
additive term is shared across documents and the multiplicative term is
positive, so each token-pair score is an increasing affine function of
the residual score; maxima and MaxSim sums preserve the order. The
invariance is exact for a shared additive mean \textbf{without}
re-normalization --- the residuals' orthogonality to $u$ is what removes
the cross-terms --- while pipelines that center and then re-normalize
perturb per-document norms and are not exactly rank-preserving; on real
embeddings this cost is measurable (Sect.~\ref{sec:calculus}) and is the
fidelity axis of the trade-off calculus.

\begin{lemma}[angle contraction]\label{lem:contraction}
In the $\rho$-cone model, any two vectors $x_i, x_j$ whose residuals
subtend the angle $\theta_{\mathrm{res}}$ subtend a full-space angle
$\theta_{\mathrm{full}}$ satisfying
$\sin(\theta_{\mathrm{full}}/2)=\sqrt{1-\mathrm{MCS}}\,\sin(\theta_{\mathrm{res}}/2)$.
\end{lemma}
\smallskip\noindent\emph{Sketch.} The shared component cancels in differences:
$\|x_i-x_j\|^2=(1-\rho^2)\|z_i-z_j\|^2$. For unit vectors
$\|a-b\|=2\sin(\theta/2)$, and $\mathrm{MCS}=\rho^2$ in the model;
the identity follows. The contraction is exact for $\sin(\theta/2)$;
for SimHash it bounds the Gaussian-hyperplane separation probability
$\theta_{\mathrm{full}}/\pi$ \cite{goemans1995maxcut} (used in
Theorem~\ref{thm:degeneration}(i)).

\begin{remark}[mixed cone weights]
Real embeddings carry per-vector weights rather
than a shared $\rho$; the identity generalizes to
$\cos\theta_{\mathrm{full}}=\rho_q\rho_d+\sqrt{(1-\rho_q^2)(1-\rho_d^2)}\,\cos\theta_{\mathrm{res}}$,
and the results degrade gracefully ($O(\delta)$ slack for deviations of norm at
most $\delta$ from the shared-$\rho$ ideal).
\end{remark}

\begin{theorem}[anisotropy degeneration of FDEs]\label{thm:degeneration}
Let $Q$ be a query's token set and $D^\star, D'$ two documents' token
sets in the $\rho$-cone model, and let $m$ be the total number of tokens
across the three sets. Consider a MUVERA-style FDE with
$k_{\mathrm{sim}}$ SimHash hyperplanes per repetition and $R$
repetitions. (i) With probability at least
$1-m^2 k_{\mathrm{sim}}R\sqrt{1-\mathrm{MCS}}$, every token of
$Q\cup D^\star\cup D'$ lands in the same bucket in every repetition.
(ii) Conditional on this event, each repetition's FDE score equals the
mean-pooled score up to a fixed aggregation factor ($m_q$ under
MUVERA's aggregation; $m_q m_d$ under sum pooling) --- exactly without
the inner projection, in expectation with it. (iii) For any $Q$ whose
residual mean $\bar z$ satisfies $0<\|\bar z\|<1$, there exist
$D^\star, D'$ for which exact MaxSim ranks $D^\star$ strictly above
$D'$ by a margin $(1-\mathrm{MCS})\Delta$, for an instance-dependent
$\Delta>0$, while the collapsed FDE ranks $D'$ above $D^\star$.
\end{theorem}
\smallskip\noindent\emph{Sketch.} (i) By Lemma~\ref{lem:contraction}, a fixed hyperplane
separates a pair with probability
$\theta_{\mathrm{full}}/\pi\le\sin(\theta_{\mathrm{full}}/2)\le\sqrt{1-\mathrm{MCS}}$
(the first inequality because $\sin(\theta/2)$ is concave on $[0,\pi]$
and meets its chord $\theta/\pi$ at both endpoints; the second by
Lemma~\ref{lem:contraction} with $\sin\le 1$). A union bound over the at
most $m^2$ query--document cross-pairs, $k_{\mathrm{sim}}$ hyperplanes,
and $R$ repetitions gives the event; same-side of a hyperplane is
transitive, so cross-pair collision implies full collision. (ii) Under
full collision, for sum pooling the score is exactly
$\sum_i\sum_j\langle q_i,d_j\rangle=m_q m_d\,\langle\operatorname{mean}(Q),\operatorname{mean}(D)\rangle$,
with $m_q, m_d$ the query and document token counts (exactly before the
inner Rademacher projection; with the projection, in expectation) ---
mean pooling, which discards the max structure. (iii) Choose $D'$ as
near-copies of $\operatorname{mean}(Q)$'s residual direction and
$D^\star$ as the exact residual match of each $q_i$ individually but
mutually diverse, both with equal token counts so the $m_d$ factor
cancels; the mean-pool score of $D'$ then exceeds $D^\star$'s, while
$\mathrm{MaxSim}(D^\star)-\mathrm{MaxSim}(D')=(1-\mathrm{MCS})\Delta$
with $\Delta$ the corresponding residual-space MaxSim gap. Exact
scoring stays correct by Lemma~\ref{lem:invariance} while the FDE
inverts the ranking. Under MUVERA's actual asymmetric aggregation
(document side = bucket mean, query side = sum) the collapse is cleaner
still --- the score becomes
$m_q\langle\operatorname{mean}(Q),\operatorname{mean}(D)\rangle$,
with no document-dependent scale --- and the same instance inverts.

\begin{remark}[regime]
The probability bound in (i) engages only as
$\mathrm{MCS} \to 1$: the union bound over $m^2 k_{\mathrm{sim}} R$
pairs is extremely loose, and at the operating points measured in
Sect.~\ref{sec:fde} (token MCS $\approx 0.9$) the all-pairs event it
certifies is rare even though measured collision rates are high. The
theorem certifies the mechanism and the limit --- sufficient anisotropy
forces collision and degeneration --- not a quantitative account of
collision rates at measured MCS; those are Sect.~\ref{sec:fde}'s
measurements.
\end{remark}

\begin{corollary}[error-side inflation of the data-oblivious guarantee]\label{cor:inflation}
Let an FDE configuration satisfy MUVERA's additive guarantee
\cite[their Theorem~2.1]{dhulipala2024muvera}: the FDE inner
product approximates the normalized Chamfer similarity
$\mathrm{NChamfer}(Q,D)=\frac{1}{|Q|}\sum_{q\in Q}\max_{d\in D}\langle q,d\rangle$
within $\pm\varepsilon$ for arbitrary unit-vector sets. In the
$\rho$-cone model, the same configuration certifies the
ranking-relevant residual scores only within
$\pm\,\varepsilon/(1-\mathrm{MCS})$.
\end{corollary}
\smallskip\noindent\emph{Sketch.} Lemma~\ref{lem:invariance} gives
$\mathrm{NChamfer}_{\mathrm{full}} = \rho^2 +
(1-\rho^2)\,\mathrm{NChamfer}_{\mathrm{res}}$ exactly, so differences
between documents' scores --- the quantities that decide a ranking ---
contract by exactly $(1-\mathrm{MCS})$; an additive-$\varepsilon$
certificate on full-space scores is therefore an additive
$\varepsilon/(1-\mathrm{MCS})$ certificate on residual scores. (Stated
budget-side instead: any guarantee parameter scaling as
$\varepsilon^{-2}$ --- the hyperplane count $k_{\mathrm{sim}}$ and the
projection dimension $d_{\mathrm{proj}}$ in their Theorem 2.1 ---
inflates by $(1-\mathrm{MCS})^{-2}$ at fixed residual-score error; the
corollary is the error-side statement.)

\begin{proposition}[translation invariance of residual schemes]\label{prop:translation}
Translating all embeddings shifts a centroid/residual scheme's
centroids but leaves every residual distance unchanged.
\end{proposition}
\smallskip\noindent\emph{Sketch.} Centroids of shifted points shift by the same vector.
Beyond the invariance, cone concentration reduces the effective
dimensionality the coarse partition must cover; whitening reverses
this by expanding low-variance directions, which can harm both the
coarse assignment and the residual code --- adaptive partitions
therefore need not want isotropy.

\begin{proposition}[hubness invariance]\label{prop:hubness}
L2 hubness computed from pairwise distances --- the skewness of the
$k$-occurrence distribution --- is invariant under centering and
orthonormal rotation.
\end{proposition}
\smallskip\noindent\emph{Sketch.} Both maps are isometries of pairwise L2 distances. The
invariance is what removes hubness as a lever for these transforms;
its empirical relationship to graph recall --- inconsistent as a
predictor --- is examined in Sect.~\ref{sec:hnsw}.

\begin{proposition}[misalignment factor]\label{prop:mf}
Partition the coordinates, in the operating basis, into $M$ blocks of
dimension $d_s = d/M$; let $\Sigma_j$ denote the $j$-th diagonal block
of the positive-definite embedding covariance $\Sigma$
(Sect.~\ref{sec:measurement}), and give every block an equal rate of
$b$ bits (structural in PQ). Then high-rate distortion satisfies
$D \propto \sum_j |\Sigma_j|^{1/d_s}$, and
\[
\mathrm{MF} \;=\; \frac{\tfrac{1}{M}\sum_j |\Sigma_j|^{1/d_s}}{|\Sigma|^{1/d}}
\;=\; [\text{AM--GM gap}]\times[\text{Fischer gap}] \;\ge\; 1,
\]
with the first factor $=1$ iff block determinants balance and the
second $=1$ iff $\Sigma$ is block-diagonal.
\end{proposition}
\smallskip\noindent\emph{Sketch.} High-rate Gaussian theory
\cite{zador1982asymptotic,gersho1979asymptotically} gives block
distortion $\propto d_s\,2^{-2b/d_s}|\Sigma_j|^{1/d_s}$, so total
distortion tracks $\sum_j |\Sigma_j|^{1/d_s}$. By AM--GM,
$\sum_j |\Sigma_j|^{1/d_s} \ge M(\prod_j|\Sigma_j|)^{1/d}$ with
equality iff block determinants balance --- the imbalance cost; in the
diagonal case this is the classical equal-rate waste on unequal
sources, and an orthonormal rotation can rebalance it without labels.
By Fischer's inequality, $\prod_j|\Sigma_j| \ge |\Sigma|$ with
equality iff the covariance is block-diagonal --- the correlation cost.
The floor $M|\Sigma|^{1/d}$ is rotation-invariant; attainability of
the floor is approximate (Ge et al.'s eigenvalue allocation
\cite{ge2013opq}), matching measured OPQ landing near $1.8$ rather
than $1$. The distortion-to-recall bridge --- $\mathrm{logit}(s)$ linear
in $\log(\sqrt{\mathrm{MF}\cdot\mathrm{floor}}/\mathrm{margin})$, with
$s$ the efficiency of Sect.~\ref{sec:axes} at the stated budget and
margin the score margin of Sect.~\ref{sec:measurement} --- is derived
under a Gaussian noise-versus-margin survival argument and validated
by calibration (Sect.~\ref{sec:evaluation}). The evaluated
configuration ($8$ bits per block) sits below the high-rate regime, so
the bridge treats $\sum_j|\Sigma_j|^{1/d_s}$ as a proxy whose adequacy
is established by that calibration, not by the asymptotics.

\begin{proposition}[fidelity gap model; Sect.~\ref{sec:fidelity}]\label{prop:gap}
For an item of the base (untransformed) exact top-$k$ whose
exact-score gap to the depth-$N$ boundary is $\gamma_i$ ($k$, $N$ as in
Sect.~\ref{sec:axes}), and a transform whose gap perturbations ---
measured on (query, item, boundary-item) triples --- are Gaussian with
measured mean shift and standard deviation $\sigma$, the survival
probability is $\Phi((\gamma_i+\mathrm{shift})/\sigma)$, with $\Phi$
the standard normal CDF.
\end{proposition}
\smallskip\noindent\emph{Sketch.} A transform evicts the item only if the perturbation of the
score \emph{gap} exceeds $\gamma_i$. Near neighbors perturb coherently,
which is why independent-perturbation models over-predict fidelity
loss several-fold; the model's known failure mode (optimistic at
extreme distortion) stems from using the native boundary item as
proxy.

\begin{proposition}[IVF routing, functional form; Sect.~\ref{sec:ivf}]\label{prop:routing}
Writing a query as its neighbor plus a perturbation, the probability
that the neighbor's cell leaves the query's nprobe highest-ranked
cells is a gap-survival event over centroid-distance gaps, with scale
set by the routing margin (Sect.~\ref{sec:measurement}); hence
$\log(\text{routing margin})$ carries a positive partial effect on
$\mathrm{logit}(s)$, with $s$ the Sect.~\ref{sec:axes} efficiency.
\end{proposition}
\smallskip\noindent\emph{Sketch.} In-cell ranking adds the score-margin term; validity of the
cell model scales with cluster excess. Constants are fitted;
coefficient-level interpretation uses marginal analyses (the joint fit
is collinearity-unstable; Sect.~\ref{sec:ivf}).

\section{Experimental Details}\label{app:details}

\paragraph{Corpora and encoders.} Dense: MS MARCO passage shards (in-sample 50k;
held-out 25k with document and query offsets past all in-sample data,
qrel-relevant documents pulled into shard) and BEIR SciFact, NFCorpus,
SciDocs (full corpora; test-split queries). Encoders: 17 sentence-transformers
models spanning 384/768/1024 dimensions (BGE small/base/large,
E5 small/base/large, MiniLM L6/L12, MPNet, GTE base/large, GTE-ModernBERT,
Nomic v1.5, MXBAI-large, Arctic m/l, UAE-Large, GIST), per-model
query/document prompts per each model card. Late-interaction: GTE-ModernColBERT,
ColBERTv2, answerai-colbert-small, Jina-ColBERT-v2 via PyLate; 25k-passage
MS MARCO shard and SciFact; token-level ragged storage.

\paragraph{Indexes.} faiss: PQ ($M{=}16$, 8-bit), OPQ, IVF (nlist 1024, nprobe 16;
budget grids where stated), HNSW ($M{=}32$, efConstruction 100, efSearch
grids 16--128). FDE: vendored MUVERA (SimHash $k_{sim}{=}5$, $R{=}8$,
projection dim 16; document = bucket mean with empty-bucket fill, query =
sum). Candidate depth $N{=}100$, $k{=}10$.

\paragraph{Measurement suite} (identical for every dataset--transform--index run):
mean/covariance resolutions incl.\ the misalignment factor; score margins;
fidelity at depths 10/50/100; efficiency (self-recall) on budget grids;
cluster panel ($k{=}64$ on a 10k sample; twin-normalized excess versions);
local-structure panel; qrel recall where labels exist (\texttt{msmarco\_heldout}:
300 judged queries, 315 relevance pairs).

\paragraph{Probe training (Sect.~\ref{sec:learning}).} Dense probes (frozen encoder, label-free):
InfoNCE on exact-neighbor pseudo-positives as the rank term, plus the
variant's geometry term --- none (rank-only), an orthogonality-projected
rotation (dose-corrected at $\lambda{=}300$, $1{,}500$ steps), or the
navigability objective (residual MLP; each document's distance to its
own base-space centroid against the nearest other; doses $\lambda{=}10,
50$) --- with an alignment term anchoring each embedding to its original.
Token probe (FDE): per-token affine map; cone term = squared norm of
the mean transformed token (the differentiable form of token MCS);
alignment term = mean $(1-\cos)$ to the original token; cone weights
$1, 4, 16$.

\paragraph{Synthetic causal sweeps (Sect.~\ref{sec:fde}).} Token bags on the unit sphere,
built from a shared vocabulary of isotropically drawn unit ``kernel''
directions; anisotropy is dialed by mixing each token with a common
unit direction on an appended axis, $z \to (\sqrt{1-\rho^2}\,z,\;
\rho)$, so MCS $= \rho^2$ while every inner product undergoes the same
affine map and exact rankings are unchanged. Coarse relevance: each
document and query draws its tokens from a random subset of kernels, so
relevance is topic-set overlap and survives mean pooling. Fine-grained
relevance: $10\%$ of each document's tokens come from one
document-specific signal kernel and the rest from a small shared
background pool, each query probing a single signal kernel --- exact
MaxSim resolves the minority signal, mean pooling drowns it. Five seeds
per grid point.

\paragraph{Predictor-comparison detail (Sect.~\ref{sec:twin25k}, Table~\ref{tab:fullcorpus}).}
Per-fraction naive subsample readouts (MAE, HNSW then
IVF): 2\% (513 docs) $0.090$, $0.496$; 5\% ($1{,}283$) $0.072$, $0.531$;
10\% ($2{,}566$) $0.055$, $0.409$; 20\% ($5{,}131$) $0.038$, $0.342$ ---
the extrapolation beats every fraction it is fitted from. Query-source
ladder on identical full-size twin corpora (MAE, HNSW then IVF): ridged
Gaussian on the $1{,}000$ real queries $0.207$, $0.311$; cluster-traffic
mixture ($64$ weights, $+0.5$ smoothing) $0.060$, $0.073$;
document-proxy (default recipe) $0.040$, $0.033$; real queries
$0.012$, $0.027$. The ladder rows are three-seed re-mints of the
full-size twin; the default single-draw instrument (Table~\ref{tab:fullcorpus}) reads
$0.039$/$0.034$ for the document-proxy configuration, within its stated
draw agreement. Query traffic touches $63$ of $64$ clusters; the
fringe-offset scalar (mean cosine to own centroid: queries $0.659$,
documents $0.708$) is the measured within-cluster mismatch.

\paragraph{Operational costs (recorded; single Apple M3 Max workstation;
diagnostic single-run costs, not serving benchmarks).} Statistics pass
(token MCS rule): $0.27$--$0.55$ s per encoder once embeddings exist.
Fitted dense panels including exhaustive top-$100$: $12.5$--$15.2$ s.
Subsampled-real-index baseline, full grid: $125$--$147$ s per encoder.
Full $1$M target builds: IVF $0.9$--$1.4$ s, HNSW $131$--$157$ s; process
peaks $6.8$--$12$ GB. Twin pipeline at $1$M (mixture fit, full-size
sampling, simulation): tens of minutes per encoder on the same machine.

\paragraph{Licensing.} jina-embeddings-v3 and Jina-ColBERT-v2 are CC BY-NC 4.0
(noncommercial research evaluation); HotpotQA/BEIR HotpotQA is CC BY-SA
4.0; the remaining encoder licenses are recorded in the artifact
manifests.

\paragraph{Twins.} Tier 1: Gaussian at the corpus $(\mu, \Sigma)$, Cholesky
sampling, unit-normalized. Tier 2: $k{=}64$ mixture with empirical weights,
centroids, and within-cluster covariances (5\% mean-eigenvalue ridge),
matched to the cluster panel within a 2\% tolerance.

\paragraph{Forecast rounds.} Round 1: 22 pairs (11 encoders $\times$ 2 datasets).
Round 2: 34 dense + 6 late-interaction fresh pairs. Per-round records
and analysis outputs are included in the artifact. Seeds: 3 per stochastic index for all
reported families; synthetic sweeps 5 seeds.

\paragraph{Compute.} Encoding on Apple MPS (minutes to $\sim$1 h per model-dataset);
all measurement on CPU. torch and faiss never share a process (macOS
libomp conflict).

\bibliographystyle{splncs04}
\bibliography{references}

\begin{thebibliography}{10}
\providecommand{\url}[1]{\texttt{#1}}
\providecommand{\urlprefix}{URL }
\providecommand{\doi}[1]{https://doi.org/#1}

\bibitem{amsaleg2015lid}
Amsaleg, L., Chelly, O., Furon, T., Girard, S., Houle, M.E., Kawarabayashi,
  K.i., Nett, M.: Estimating local intrinsic dimensionality. In: ACM SIGKDD
  International Conference on Knowledge Discovery and Data Mining (KDD) (2015)

\bibitem{aumuller2021local}
Aum{\"u}ller, M., Ceccarello, M.: The role of local dimensionality measures in
  benchmarking nearest neighbor search. Information Systems  \textbf{101}
  (2021)

\bibitem{bruch2025optimistic}
Bruch, S., Krishnan, A., Nardini, F.M.: Optimistic query routing in
  clustering-based approximate maximum inner product search. In: Advances in
  Neural Information Processing Systems (NeurIPS) (2025)

\bibitem{calinski1974dendrite}
Cali{\'n}ski, T., Harabasz, J.: A dendrite method for cluster analysis.
  Communications in Statistics  \textbf{3}(1),  1--27 (1974)

\bibitem{dhulipala2024muvera}
Dhulipala, L., Hadian, M., Jayaram, R., Lee, J., Mirrokni, V.: {MUVERA}:
  Multi-vector retrieval via fixed dimensional encodings. In: Advances in
  Neural Information Processing Systems (NeurIPS) (2024)

\bibitem{elliott2024hnsw}
Elliott, O.P., Clark, J.: The impacts of data, ordering, and intrinsic
  dimensionality on recall in hierarchical navigable small worlds. In: ACM
  SIGIR International Conference on Theory of Information Retrieval (ICTIR)
  (2024)

\bibitem{ethayarajh2019contextual}
Ethayarajh, K.: How contextual are contextualized word representations?
  comparing the geometry of {BERT}, {ELMo}, and {GPT-2} embeddings. In:
  Conference on Empirical Methods in Natural Language Processing (EMNLP) (2019)

\bibitem{gao2024rabitq}
Gao, J., Long, C.: {RaBitQ}: Quantizing high-dimensional vectors with a
  theoretical error bound for approximate nearest neighbor search. Proceedings
  of the ACM on Management of Data (SIGMOD)  (2024)

\bibitem{gao2019degeneration}
Gao, J., He, D., Tan, X., Qin, T., Wang, L., Liu, T.Y.: Representation
  degeneration problem in training natural language generation models. In:
  International Conference on Learning Representations (ICLR) (2019)

\bibitem{ge2013opq}
Ge, T., He, K., Ke, Q., Sun, J.: Optimized product quantization for approximate
  nearest neighbor search. In: IEEE Conference on Computer Vision and Pattern
  Recognition (CVPR) (2013)

\bibitem{gersho1979asymptotically}
Gersho, A.: Asymptotically optimal block quantization. IEEE Transactions on
  Information Theory  \textbf{25}(4),  373--380 (1979)

\bibitem{goemans1995maxcut}
Goemans, M.X., Williamson, D.P.: Improved approximation algorithms for maximum
  cut and satisfiability problems using semidefinite programming. Journal of
  the ACM (JACM)  \textbf{42}(6),  1115--1145 (1995)

\bibitem{guo2020scann}
Guo, R., Sun, P., Lindgren, E., Geng, Q., Simcha, D., Chern, F., Kumar, S.:
  Accelerating large-scale inference with anisotropic vector quantization. In:
  International Conference on Machine Learning (ICML) (2020)

\bibitem{hara2015localized}
Hara, K., Suzuki, I., Shimbo, M., Kobayashi, K., Fukumizu, K., Radovanovi{\'c},
  M.: Localized centering: Reducing hubness in large-sample data. In: AAAI
  Conference on Artificial Intelligence (AAAI) (2015)

\bibitem{he2012difficulty}
He, J., Kumar, S., Chang, S.F.: On the difficulty of nearest neighbor search.
  In: International Conference on Machine Learning (ICML) (2012)

\bibitem{houle2013dimensionality}
Houle, M.E.: Dimensionality, discriminability, density and distance
  distributions. In: IEEE International Conference on Data Mining Workshops
  (ICDMW) (2013)

\bibitem{indyk2023worstcase}
Indyk, P., Xu, H.: Worst-case performance of popular approximate nearest
  neighbor search implementations: Guarantees and limitations. In: Advances in
  Neural Information Processing Systems (NeurIPS) (2023)

\bibitem{jegou2011pq}
J{\'e}gou, H., Douze, M., Schmid, C.: Product quantization for nearest neighbor
  search. IEEE Transactions on Pattern Analysis and Machine Intelligence
  (TPAMI)  \textbf{33}(1),  117--128 (2011)

\bibitem{jung2023isotropic}
Jung, E., Park, J., Choi, J., Kim, S., Rhee, W.: Isotropic representation can
  improve dense retrieval. In: Pacific-Asia Conference on Knowledge Discovery
  and Data Mining (PAKDD) (2023)

\bibitem{kumar2024ehi}
Kumar, R., Mittal, A., Gupta, N., Kusupati, A., Dhillon, I., Jain, P.: {EHI}:
  End-to-end learning of hierarchical index for efficient dense retrieval.
  Transactions on Machine Learning Research (TMLR)  (2024)

\bibitem{lakshman2025stability}
Lakshman, V., Munyampirwa, B., Shun, J., Coleman, B.: Breaking the curse of
  dimensionality: On the stability of modern vector retrieval. arXiv preprint
  arXiv:2512.12458  (2025)

\bibitem{li2020bertflow}
Li, B., Zhou, H., He, J., Wang, M., Yang, Y., Li, L.: On the sentence
  embeddings from pre-trained language models. In: Conference on Empirical
  Methods in Natural Language Processing (EMNLP) (2020)

\bibitem{li2023jtr}
Li, H., Ai, Q., Zhan, J., Mao, J., Liu, Y., Liu, Z., Cao, Z.: Constructing
  tree-based index for efficient and effective dense retrieval. In:
  International ACM SIGIR Conference on Research and Development in Information
  Retrieval (SIGIR) (2023)

\bibitem{malkov2018hnsw}
Malkov, Y.A., Yashunin, D.A.: Efficient and robust approximate nearest neighbor
  search using hierarchical navigable small world graphs. vol.~42, pp. 824--836
  (2020)

\bibitem{mu2018allbut}
Mu, J., Viswanath, P.: All-but-the-top: Simple and effective postprocessing for
  word representations. In: International Conference on Learning
  Representations (ICLR) (2018)

\bibitem{prokhorenkova2020graph}
Prokhorenkova, L., Shekhovtsov, A.: Graph-based nearest neighbor search: From
  practice to theory. In: International Conference on Machine Learning (ICML)
  (2020)

\bibitem{radovanovic2010hubs}
Radovanovi{\'c}, M., Nanopoulos, A., Ivanovi{\'c}, M.: Hubs in space: Popular
  nearest neighbors in high-dimensional data. Journal of Machine Learning
  Research (JMLR)  \textbf{11},  2487--2531 (2010)

\bibitem{rousseeuw1987silhouettes}
Rousseeuw, P.J.: Silhouettes: A graphical aid to the interpretation and
  validation of cluster analysis. Journal of Computational and Applied
  Mathematics  \textbf{20},  53--65 (1987)

\bibitem{rudman2024istar}
Rudman, W., Eickhoff, C.: Stable anisotropic regularization. In: International
  Conference on Learning Representations (ICLR) (2024), arXiv:2305.19358

\bibitem{santhanam2022plaid}
Santhanam, K., Khattab, O., Potts, C., Zaharia, M.: {PLAID}: An efficient
  engine for late interaction retrieval. In: ACM International Conference on
  Information and Knowledge Management (CIKM) (2022)

\bibitem{santhanam2022colbertv2}
Santhanam, K., Khattab, O., Saad-Falcon, J., Potts, C., Zaharia, M.:
  {ColBERTv2}: Effective and efficient retrieval via lightweight late
  interaction. In: Conference of the North American Chapter of the Association
  for Computational Linguistics (NAACL) (2022)

\bibitem{su2021whitening}
Su, J., Cao, J., Liu, W., Ou, Y.: Whitening sentence representations for better
  semantics and faster retrieval. arXiv preprint arXiv:2103.15316  (2021)

\bibitem{suzuki2013centering}
Suzuki, I., Hara, K., Shimbo, M., Saerens, M., Fukumizu, K.: Centering
  similarity measures to reduce hubs. In: Conference on Empirical Methods in
  Natural Language Processing (EMNLP) (2013)

\bibitem{timkey2021rogue}
Timkey, W., van Schijndel, M.: All bark and no bite: Rogue dimensions in
  transformer language models undermine representational similarity. In:
  Conference on Empirical Methods in Natural Language Processing (EMNLP) (2021)

\bibitem{veneroso2025crisp}
Veneroso, J., Jayaram, R., Rao, J., Hern{\'a}ndez~{\'A}brego, G., Hadian, M.,
  Cer, D.: {CRISP}: Clustering multi-vector representations for denoising and
  pruning. arXiv preprint arXiv:2505.11471  (2025)

\bibitem{zador1982asymptotic}
Zador, P.L.: Asymptotic quantization error of continuous signals and the
  quantization dimension. IEEE Transactions on Information Theory
  \textbf{28}(2),  139--149 (1982)

\bibitem{zeng2025lira}
Zeng, X., Deng, L., Chen, P., Chen, X., Su, H., Zheng, K.: {LIRA}: A
  learning-based query-aware partition framework for large-scale {ANN} search.
  In: Proceedings of the ACM Web Conference (WWW) (2025)

\bibitem{zhan2021jpq}
Zhan, J., Mao, J., Liu, Y., Guo, J., Zhang, M., Ma, S.: Jointly optimizing
  query encoder and product quantization to improve retrieval performance. In:
  ACM International Conference on Information and Knowledge Management (CIKM)
  (2021)

\bibitem{zhang2021poeem}
Zhang, H., Shen, H., Qiu, Y., Jiang, Y., Wang, S., Xu, S., Xiao, Y., Long, B.,
  Yang, W.Y.: Joint learning of deep retrieval model and product quantization
  based embedding index. In: International ACM SIGIR Conference on Research and
  Development in Information Retrieval (SIGIR) (2021)

\end{thebibliography}

\end{document}